\documentclass[aps,prb,twocolumn,groupedaddress,showpacs,superscriptaddress,amssymb,amsmath]{revtex4-1}
\usepackage{graphicx}
\usepackage{tabularx}
\usepackage{color}
\usepackage{amsmath}
\usepackage{comment}
\newcommand{\be}{\begin{equation}}
\newcommand{\ee}{\end{equation}}
\newcommand{\bea}{\begin{eqnarray}}
\newcommand{\eea}{\end{eqnarray}}
\usepackage{dcolumn}
\usepackage{hyperref}
\usepackage{bm}
\usepackage{epsf}
\usepackage{subfigure}
\usepackage{epstopdf}%
\usepackage{amsfonts}

\begin{document}

%\title{On the definition and simulations of heat transport: The good and bad definitions}
\title{On the definitions and simulations of vibrational heat transport in nanojunctions}

% demons of spurious effects
% The good and bad of heat current 
% On the good and bad def of heat current 
% On the 
%  On the defi of heat current in classical simulations: Spurious effects
% 
% Spurious 
\author{Na'im Kalantar}
\affiliation{Chemical Physics Theory Group, Department of Chemistry, 
University of Toronto, 80 Saint George St., Toronto, Ontario, Canada M5S 3H6}

\author{Bijay Kumar Agarwalla}
\affiliation{Department of Physics, Dr. Homi Bhabha Road, Indian Institute of Science Education and Research, Pune, India 411008}

\author{Dvira Segal}
\email{dvira.segal@utoronto.ca}
\affiliation{Chemical Physics Theory Group, Department of Chemistry and Centre for Quantum Information and Quantum Control,
University of Toronto, 80 Saint George St., Toronto, Ontario, Canada M5S 3H6}
\affiliation{Department of Physics, University of Toronto, Toronto, Ontario, Canada M5S 1A7}

\date{\today}

\begin{abstract}
Thermal transport through nanosystems is central to numerous processes in 
chemistry, material sciences, electrical and mechanical engineering, with classical molecular dynamics as the key simulation tool.
Here we focus on thermal junctions with a molecule bridging two solids that are maintained at different temperatures. The  classical 
steady state heat current in this system can be simulated in different ways, either at the interfaces with the solids, which are represented by thermostats, or between atoms within the conducting molecule. We show that while the latter, intramolecular definition feasibly converges to the correct limit, the molecule-thermostat interface definition is more challenging to converge to the correct result. 
The problem with the interface definition is demonstrated by simulating heat transport in harmonic and anharmonic one-dimensional chains illustrating unphysical effects such as thermal rectification in harmonic junctions.
\end{abstract}

\maketitle 

%=========================
\section{Introduction}
\label{Sec-intro}

Classical simulations of heat transfer through networks of beads and springs had played a crucial role in the development of nonlinear science \cite{FPU1,FPU2}, leading e.g. to the discovery of integrable systems with solitons as a prime example. Fundamentally, molecular dynamics simulations of vibrational heat flow in molecules aid in understanding chemical reactivity and e.g. protein folding dynamics  \cite{LeitnerRev}. For applications, understanding thermal energy propagation, redistribution, and dissipation is essential for developing electronic, mechanical, thermal and thermoelectric devices, specifically organic-inorganic heterostructures \cite{PopRev,BaowenRev,CahillRev,ReddyRev,DharRev,BijayGRev,BijayRev}.

Chains of beads and springs coupled at the edges to solids that are maintained at fixed temperatures serve to model heat transport in low dimensional systems. Abundance of simulations have demonstrated rich and often {\it anomalous} heat transport phenomena in low dimensional systems, specifically in one-dimensional (1D) chains, compared to the behavior of macroscopic objects \cite{Lebowitz,Lepri97,DharRev,Livi}. 
 Recent experiments probed the flow of vibrational energy (heat) in self-assembled monolayers of alkanes \cite{Braun,Gotsmann,Malen}, down to the single molecule junction \cite{Brendt,ReddyE}.
As well, experiments in solutions were performed based on pump-probe spectroscopy methods \cite{Rubtsov}.
 In junction experiments, the setup constitutes a linear (quasi 1D) molecule bridging two solids with the steady state  heat current, or the thermal conductance as observables of interest. 

\begin{figure} 
\includegraphics[scale=0.45]{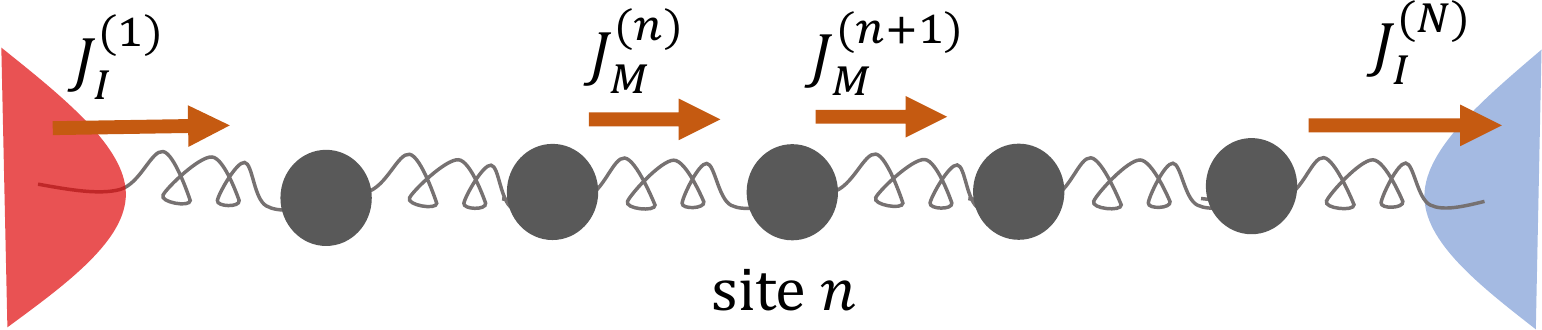}%{schemeN-crop.pdf}
\caption{
Scheme of a vibrational heat transport junction.
A molecular chain is attached to heat baths (solids) at the boundaries, with
the baths emulated by Langevin thermostats. 
The intramolecular currents $J_M^{(n)}$ are calculated from Eq. (\ref{eq:JB}).
The interface currents (at the left and right contacts) are identified by $J_I^{(1)}$ and $J_I^{(N)}$, and are 
computed from Eq. (\ref{eq:JI}). 
These expressions can be readily generalized beyond one-dimension to treat more complex structures.
}
\label{Fig1}
\end{figure}

% HERE
In this work, we study the nonequilibrium steady state vibrational heat transport in 1D chains connected to solids at the boundaries, as depicted in Fig.~\ref{Fig1}. 
The solid contacts are emulated with thermostats; here we adopt Langevin baths. 
Such classical simulations have a long and rich history \cite{LepriRev,DharRev,Livi}.
Recent studies (i) continue to address the fundamental anomalous properties of heat conduction in periodic and disordered one-dimensional chains \cite{Leitner16,Leitner18,Zwolak,WangNJP,Imry1,Imry2,CoexFPU}, (ii) recreate experimental setups to reveal transport mechanisms \cite{Inon1,Inon2,Roya},
and  (iii) provide guidelines for enhancing or suppressing the thermal conductance (oftentimes based on the harmonic force-field and using quantum scattering methods)  \cite{Pauly,Gemma,Lambert,Cuniberti19,Cuniberti19S}.
Naturally, one wonders: What problems in this field remain unresolved?

Here, we address a fundamental computational problem: How to efficiently and accurately simulate phononic heat transport in nonequilibrium situations as depicted in Fig.~\ref{Fig1}. This issue is far from being merely technical, since, as we show here, simulations that are only seemingly converged lead to faulty predictions of nonlinear functionality, and the violation of fundamental symmetries. This challenge, of converging simulations at low cost, boils down to making an adequate choice for the working definition of heat current in the system. 
In what follows we only focus on the contribution of the nuclei to thermal transport (referred to as vibrational or phononic conduction), 
 and do not consider the additional electronic contribution.

%====================================
%  Figure 1 

%======================================

% def
We introduce two {\it equivalent}, intuitive definitions for the steady state heat current in junctions:  In the intra-molecular  ($J_M$) definition, the heat current is evaluated between particles {\it within} the conducting structure. In contrast, the interface ($J_I$) heat current is defined at the molecule-solid boundary, by calculating the net heat exchange between the solid (thermostat) and the molecule. 
While these two definitions are {\it mathematically} equivalent, we show that {\it computationally} the former (intramolecular expression) is superior: The interface definition suffers from a convergence problem, related to the challenge of achieving local energy equipartition at   thermal equilibrium. In particular, the interface definition may lead to the erroneous identification of nonlinear phenomena in thermal transport, such as the (incorrect) observation of thermal rectification in harmonic chains.

It is quite interesting to note that contrasting the numerical problem,
 the interface definition is in fact advantageous in {\it formal-analytic} studies of 
anharmonic heat flow, full counting statistics, quantum heat transport, and more broadly quantum thermodynamics.
For example, consider the problem of classical or quantum heat flow
through an anharmonic structure. While the particles within the chain interact based on an anharmonic force field, the coupling between the molecular system and the bath is typically taken as a bilinear (harmonic) coupling form.
As such, it is convenient to develop a formal expression for heat exchange {\it at the contact region}, rather than within the molecule---where the current needs to be defined in terms of high-order correlation functions  \cite{BijayGRev}.
Similarly, a full-counting statistics analysis of heat exchange, which follows the two-time measurement protocol is conveniently performed by considering the difference in energy content at the baths attached to the system \cite{Bijay12}.
% As such, the problem reduces to the analysis of heat flow from the reservoirs towards the system  

As an additional example for the utility of the interface definition in formal methodologies, 
recall studies of quantum heat flow between a quantum system and a thermal bath. Given the ambiguity of a `heat current operator' \cite{Lianao}, quantum heat exchange is calculated at the contact region:
At weak system-bath coupling the time derivative of the expectation value, $d\langle H_S\rangle/dt $, 
yields the formal heat current expression
  ${\rm Tr}[H_SD_K(\rho_S)]$ with $D_K(\rho_S)$ as  the dissipator part of the dynamics due to the $K$th bath, which is responsible for heat exchange between the system and the $K$th reservoir; $\rho_S$  is the density matrix of the system and $H_S$ is the molecular (`system') Hamiltonian.  The interface definition is indeed commonly employed in studies of quantum thermodynamics, even when approaching the strong system-bath coupling limit; some examples include Refs \cite{Tanimura,Hava}.

The interface and intramolecular definitions thus nicely complement: The interface definition is advantageous in formal studies of heat exchange, particularly in anharmonic systems. In contrast, in this paper we demonstrate that the intramolecular expression shows a better performance in numerical simulations, both in the frequency and time domain.

% DDD new
Interestingly, several recent studies revisited the definition of vibrational heat flux in low dimensional systems:
In Ref. \cite{Carlo}, the heat current was represented in two different ways following either a Lagrangian or an Eulerian approach, resulting in different microscopic definitions---that showed equivalent simulation results. In Ref.  \cite{Li19}, the impact of the thermostats (Langevin vs. Nose-Hoover) on thermal conduction was analyzed, and it was found that the Nose-Hoover approach lead to incorrect thermal rectification values. Our work further contributes to this endeavour, by comparing the intramolecular and interface definitions for the vibrational heat flux.

The paper is organized as follows. 
In Sec. \ref{Sec-M}, we present the nanojunction and the Langevin equation of motion, as well as the intramolecular and interface definitions for phononic heat transport.
In Sec. \ref{Sec-H}, we focus on harmonic systems and present simulations  in the frequency domain based on the Green's function formalism.
Molecular dynamics simulation are performed in Sec. \ref{Sec-MD}, and we demonstrate the behavior of 
the different currents in both harmonic and anharmonic chains.
We conclude in Sec. \ref{Sec-Summ}.

%================================================

\section{Model and Method}
\label{Sec-M}

\subsection{Langevin thermostats}

We focus on one-dimensional molecular junctions as illustrated in Fig.~\ref{Fig1}.
We write down the classical Hamiltonian and the corresponding classical equations of motion (EOM); a quantum description based on Heisenberg EOM directly follows \cite{DharRev}, but we do not describe it here. The classical Hamiltonian is
\bea
H&=&\sum_{n=1}^N \frac{p_n^2}{2m_n}  + \sum_{n=2}^{N} V(x_{n}-x_{n-1}) 
\nonumber\\
&+& V_B(x_{N+1}-x_{N}-a) + V_B(x_{1}-x_{0}-a). 
\label{Hamiltonian}
\eea
Here, $x_n$ and $p_n$ are 
the displacements and momenta, respectively, of the $n$th bead; $x_0$ and $x_{N+1}$ are fixed points setting the boundaries.
$V$ is the intramolecular potential energy, and $V_B$ (which could have the same functional form as $V$) is the potential energy between the atoms at the boundaries and the solids,  setting the maximal extension of the molecule to $d=x_{N+1}-x_0$.
Since $x_0$ and $x_{N+1}$ are fixed, the potential energy $V_B$ acts to confine sites 1 and $N$, respectively.
One can generalize the model to include additional confining potential energy for all atoms.

To capture the solids at the left and right ends of the molecule, we further assume that the beads at the edges ($1$ and $N$) 
are coupled to independent heat baths (thermostats). 
This coupling is incorporated in the Langevin equation with a friction constants $\gamma_n$ and stochastic forces $\xi_n(t)$; 
these terms are related through the fluctuation-dissipation relation. 
The beads can represent atoms or coarse-grained units in a more complex system such as DNA. % DDD refs
The classical EOMs  for the displacements are
\bea
m_1 \ddot x_1 &=& 
-\frac{\partial V}{\partial x_1}  -\frac{\partial V_B}{\partial x_1}- \gamma_1 p_1(t) + \xi_1(t),
\nonumber\\
m_n \ddot x_n &=& 
-\frac{\partial V}{\partial x_n} ,\,\,\,\,\, n=2,3,...N-1,
\nonumber\\
m_N \ddot x_N &=& 
-\frac{\partial V}{\partial x_N}  -\frac{\partial V_B}{\partial x_N}- \gamma_N p_N(t) + \xi_N(t).
\label{eq:EOM}
\eea
The thermal-white noise is local and obeys the fluctuation-dissipation relation, 
$\langle \xi_{p}(t)\xi_{l}(t')\rangle = 2 k_BT_p\gamma_p m_p\delta (t-t') \delta_{p,l}$.
Here, $k_B$ is the Boltzmann's constant. 
%Here $T_n$ is the temperature of the Langevin bath attached to site $n$.
We emphasize  that $T_{1}$ and $T_N$ are not effective temperatures associated with sites $1$ or $N$; we do not make any assumptions on the notion
of a local temperature. Rather, $T_{1,N}$ are the temperatures of the Langevin baths attached to these sites.
%
%In Eq.  (\ref{eq:EOM}), in principle every particle is allowed to exchange energy with a local independent bath.
%In the junction configuration of Fig.~\ref{Fig1} and in simulations below only $\gamma_1\neq0$ and $\gamma_N\neq 0$, with all other friction coefficients  set to zero.

Specifically below we focus on harmonic chains with nearest-neighbor couplings,
\bea
V (x_n-x_{n-1})= \frac{1}{2}  k_{n,n-1}\left(x_n - x_{n-1} - a \right)^2.
\eea 
The same form is assumed at the contacts for $V_B$. The EOM for e.g. the $N$th site then becomes
\bea
m_N \ddot x_N &=& -k_{N,N-1}\left [x_N(t)-x_{N-1}(t)-a\right]
\nonumber\\
&+& k_{N,N+1} \left[x_{N+1}(t)-x_N(t) - a\right]- \gamma_N p_N(t) + \xi_N(t),
\nonumber\\
\label{eq:EOMhar}
\eea
where $a$ is the equilibrium distance between neighboring particles.

%As mentioned above, for generality, every bead is attached to a Langevin bath, and the interface current concerns heat exchange at the particle-bath contact.
%If only $\gamma_1$ and $\gamma_N$ are nonzero, the interface currents are the heat input to the first site
%and heat leakage at the last one, $-J_I^{(1)}=J_I^{(N)}$. The bulk current $J_B^{(n)}$ can be evaluated along the junction at every site $n$. 

%==========
\subsection{Heat current definitions}

Here, we derive the intramolecular and interface definitions of heat currents
as used in Langevin molecular dynamics simulations \cite{LepriRev,DharRev}.
We focus on the steady state limit---once the effect of the initial conditions subside. 
Employing the EOM (\ref{eq:EOM}) for site $N$:
\bea
&&\langle p_N(t) \dot p_N(t) \rangle =  \frac{1}{2}\frac{d}{dt} \langle p_N(t)^2\rangle 
\nonumber\\
&&=
-\left\langle \frac{\partial V}{\partial x_N} p_N(t)\right\rangle 
-\left\langle \frac{\partial V_B}{\partial x_N} p_N(t)\right\rangle
\nonumber\\
&&
- \gamma_N \left\langle p_N(t)^2\right\rangle + \langle \xi_N(t)p_N(t)\rangle.
\eea
In steady state, the local-site energy  is constant. Here, the  local energy at site $N$ is the sum of kinetic energy plus the $V_B$ confining potential.   
Using $\frac{d}{dt}\left(\frac{\langle p_N(t)^2\rangle}{2m_N}\right)+
\left\langle \frac{\partial V_B}{\partial x_N} v_N(t)\right\rangle =0$,  we get
\bea
-\left\langle \frac{\partial V}{\partial x_N} v_N(t)\right\rangle = \frac{\gamma_N}{m_N} \left\langle p_N(t)^2\right\rangle - \frac{1}{m_N}\langle \xi_N(t)p_N(t)\rangle.
\nonumber\\
\label{eq:defJ}
\eea
The average is done over initial conditions and by calculating heat exchange over a long time interval.
$v_n=p_n/m_n$ is the velocity of the $n$th particle.

We identify the left hand side of Eq. (\ref{eq:defJ}) as the intramolecular current, $J_M^{(N)}$, flowing between site $N-1$ and site $N$. 
Since in steady state, $J_M^{(n)} = J_M^{(n+1)}$ we generally define the intramolecular current at site $n$ as
\bea
J_M^{(n)} = \langle v_n  F_{n}\rangle.
\label{eq:JB}
\eea
Here, $F_n=-V'(x_{n}-x_{n-1})$ is the force exerted from the $(n-1)$th particle on the $n$th bead; derivative is taken with respect to the argument $x_n - x_{n-1}$.

Since $\langle  v_n F_n\rangle = \langle v_n F_{n+1}\rangle$,
It is common to calculate
the intramolecular current based on an averaged two-bead expression by using either one of these expressions ($1<n<N$),
$J_M^{(n)} = \frac{1}{2} \langle v_n \left( F_{n+1}+F_n\right)\rangle  =\frac{1}{2} \langle \left( v_n + v_{n-1}\right) F_n\rangle$.

Next, we identify the interface currents;  $J_I^{(N)}$ is given by the right hand side of Eq. (\ref{eq:defJ}). By making use of the
relationship $\langle p_n(t) \xi_l(t) \rangle = \gamma_n m_n k_BT_n\delta_{n,l}$
\cite{Muga} we get % DDD factor of 2 in simulations? Na'im mentioned that. 
%$n$th contact as the net heat exchange between the Langevin bath and the attached atom,
\bea
J_I^{(N)}&=& \gamma_N\left(  \frac{\langle p_N^2\rangle}{m_N}  -k_BT_N \right),
\nonumber\\
J_I^{(1)}&=&- \gamma_1\left(  \frac{\langle p_1^2\rangle}{m_1}  -k_BT_1 \right),
\label{eq:JI}
\eea
where similar considerations as in Eq. (\ref{eq:defJ}) resulted in $J_I^{(1)}$. 
The interface currents can be interpreted as the net heat exchange between the particles (1, $N$) and the attached thermostats, which are maintained at temperatures $T_{1,N}$. Note that the physical dimension of $\gamma$ is inverse time.
Particularly, at thermal equilibrium we expect that the interface current would vanish based on the principle of energy equipartition.

%As mentioned above, for generality so far in our expressions we allowed every bead to be attached to an independent 
%thermostat;  the interface current concerns heat exchange at the particle-bath contact.
%If only $\gamma_1$ and $\gamma_N$ are nonzero, the interface currents concern the heat input to the first site
%and heat drain at the last site, $-J_I^{(1)}=J_I^{(N)}$. 
In our sign convention, positive current flows from left (site 1) to right (site $N$).
The intramolecular  current $J_M^{(n)}$ can be evaluated along the junction between every two sites, and we usually study it at the center of the chain. For simplicity, in what follows we set masses to one, $m_n=1$.

% advantage of inteface def
The heat current definitions, Eqs.  (\ref{eq:JB}) and (\ref{eq:JI}) are obviously equivalent: 
They are related based on the assumption of a white Gaussian noise. 
As we mentioned in the Introduction, the interface definition is appealing in formal 
calculations: It allows the construction of a closed-form definition for the heat current even in anharmonic models, as long as the system-bath coupling is bilinear. 
Furthermore, it bypasses the problem of defining a heat current operator for quantum systems, which is non-unique \cite{Lianao}.
The interface expression is also appealing if the conducting system has a complex connectivity with many bonds
(say beyond nearest neighbors) contributing to heat propagation. In such a tangled scenario, it seems more feasible to calculate the steady state heat current by measuring the input or output power---between the system and the thermostats. In fact, in experimental studies %of heat transport through molecular junctions 
the molecular thermal conductance is evaluated in this manner, by measuring the input heat power at the contact region \cite{ReddyE}. The interface definition is also useful in hybrid models: In a recent study of thermal transport across a metal-polymer interface, heat exchange between the electronic and nuclear degrees of freedom was inferred from the net vibrational heat transfer, from the atoms to the Langevin thermostat, that is, using the interface definition \cite{Coker}.

The intramolecular definition is attractive in molecular dynamics simulations: Since in steady state the time-averaged currents $J_M^{(n)}$ are equal for every $n$,   one can perform an additional averaging, $\bar J_M\equiv \sum_{2}^{N} J_M^{(n)}/(N-1)$, to reduce the error of the individual bond current. This type of averaging can be also implemented  in complex networks.
  
 %While the interface definition of heat current is appealing mostly for formal work, 
In what follows, we examine the interface and `bare' intramolecular definitions (without site averaging), 
Eqs.  (\ref{eq:JB}) and (\ref{eq:JI})  in classical systems 
based on numerical simulations for both
harmonic and anharmonic junctions as depicted in Fig.~\ref{Fig1}. 
We show that it is more challenging  to  converge $J_I$ to the correct result---compared to  $J_M$. For example, at thermal equilibrium
$J_M$ more feasibly approaches the correct behavior (zero current) even when equipartition of energy is not yet accomplished (due to numerical errors). Away from equilibrium, $J_I$ may display a thermal diode effect for harmonic systems, which is a numerical artifact emerging from the finite time step error or the frequency integration error.

%=======================================================
\vspace{4mm}
\section{Frequency domain: Numerical integration}
\label{Sec-H}

In this Section we focus on harmonic systems and illustrate the flaws of the boundary definition $J_I$ when
working in Fourier's - frequency space. 
In the frequency domain,  the EOM (\ref{eq:EOMhar}) can be solved as an algebraic problem in steady state, and the different correlation functions (position-position, velocity-position, velocity-velocity) can be obtained analytically.

To simplify notation, in the following derivations we formally include a local thermal bath at each site, therefore introduce the friction coefficients $\gamma_{n}$ for every site, $n=1,2,..N$. In simulations we set $\gamma_{2,3,..N-1}=0$ and recover the junction setup with only two heat baths at the boundaries.
 
 \subsection{Discussion of heat current definitions}
For harmonic systems, the intramolecular heat current evaluated between sites $n$ and $n+1$---also referred to as bond current--- is given by  \cite{Roy06,DharRev} (masses are set to unit),
% % added k_{n,n+1}
\bea
&&J_M^{(n+1)}=
\nonumber\\
&&-k_{n,n+1}\sum_{m=1,N} \gamma_m k_BT_m
 \int_{-\infty}^{\infty} d\omega \frac{\omega}{\pi} {\rm Im}\left[
({\rm G}^r(\omega))_{n,m}   ({\rm G}^{a}(\omega))_{m,n+1}  
\right].
\nonumber\\
\label{eq:JBHA}
\eea
%
%\bea
%J_B^{(n)}=\sum_m \gamma_n\gamma_m \int_{-\infty}^{\infty} d\omega \frac{\omega^2}{\pi} 
%\left|({\rm G}^r(\omega))_{n,m}\right|^2k_B(T_n-T_m).
%\nonumber\\
%\label{eq:JBH}
%\eea
%
Here,  $\boldsymbol{{\rm G}^r}(\omega)$ is a symmetric matrix, corresponding to the retarded Green's function \cite{DharRev}.
 For example, for a 3-site chain with nearest neighbor interaction we write ($k_{i,j}=k_{j,i})$,
\begin{widetext}
\begin{equation*}
\label{eq:GG}
\boldsymbol{{\rm G}^r}(\omega) = \left[
\begin{array}{ccc}
-\omega^2+ k_{2,1}+k_{1,0} - i\gamma_1\omega & -k_{2,1} & 0  \\ %& -k_{4,1}\\
-k_{2,1} &   -\omega^2+ k_{2,1} + k_{3,2}- i\gamma_2 \omega   & -k_{3,2}  \\ %& -k_{4,2}    \\
0 & -k_{3,2}  &  -\omega^2+k_{3,2}+k_{4,3} - i\gamma_3\omega \\ %& -k_{3,4} \\
%-k_{4,1} & -k_{4,2}  & -k_{4,3} & -\omega^2+\sum_{p=0}^{N+1} k_{4,p} + i\gamma_2\omega  \\
\end{array}
\right]^{-1}
\end{equation*}
\end{widetext}
Furthermore,  ${\rm G}^{a}=({\rm G}^{r})^{\dagger}$ thus $[({\rm G}^{r}(\omega))_{n+1,m}]^*= ({\rm G}^a(\omega))_{m,n+1}$.
%In Appendix A we prove that this current identically vanishes at equilibrium, when $T_1=T_N$.

To calculate the interface current, we evaluate the momentum autocorrelation function  \cite{DharRev},
\bea
\langle  p_n^2 \rangle=
\frac{1}{\pi}\int_{-\infty}^{\infty} d\omega\omega^2 \sum_m \left|(G^r(\omega))_{n,m}\right|^2\gamma_mk_BT_m.
\label{eq:p2}
\eea
Note that at thermal equilibrium, with all temperatures  identical, we reach energy equipartition and we obtain the normalization condition:
\bea
N_n\equiv
\frac{1}{\pi}\int_{-\infty}^{\infty} d\omega\omega^2 \sum_m \left|(G^r(\omega))_{n,m}\right|^2\gamma_m =1.
\label{eq:N}
\eea
In Appendix A we point out that the normalization condition in fact is a strict sum rule condition for the spectral function of the chain.
Based on the definition (\ref{eq:JI}), the interface current at site 1 is given by
\bea
J_I^{(1)}&=& -\gamma_1\left(  \langle p_1^2\rangle  -k_BT_1 \right).
\nonumber\\
&=&-\sum_m\frac{\gamma_1\gamma_m}{\pi}\int_{-\infty}^{\infty} d\omega\omega^2  \left|(G^r(\omega))_{1,m} \right|^2k_BT_m +\gamma_1k_BT_1.
\nonumber\\
\label{eq:JIH}
\eea
We thus have two working definitions for the heat current: 
(i) The intramolecular current Eq. (\ref{eq:JBHA}). 
(ii)  The interface definition at site 1,
Eq. (\ref{eq:JIH}), with an analogous expression at site $N$.

In simulations, the limits of integrations are truncated at $\pm\omega_c$, and one would naively assume that taking  $\omega_c$ to be one order of magnitude larger that the thermal energy $k_BT$ and the friction coefficient $\gamma$ 
 should suffice for achieving converged results.
However, as we now show with simulations, the interface definition (\ref{eq:JIH}) is  challenging to converge  to the correct result even when $\omega_c$ is made quite high, unlike the intramolecular calculation. This problem manifests itself as a residual $J_I$ current at thermal equilibrium and as an erroneous rectification effect for $J_I$ in harmonic junctions. 

%====================================
%  Figure 2
\vspace{5mm}
\begin{figure} [htbp]
\begin{minipage}[b]{0.45\textwidth}
%\begin{subfigure}[b]{0.4\textwidth}
\includegraphics[scale=0.3]{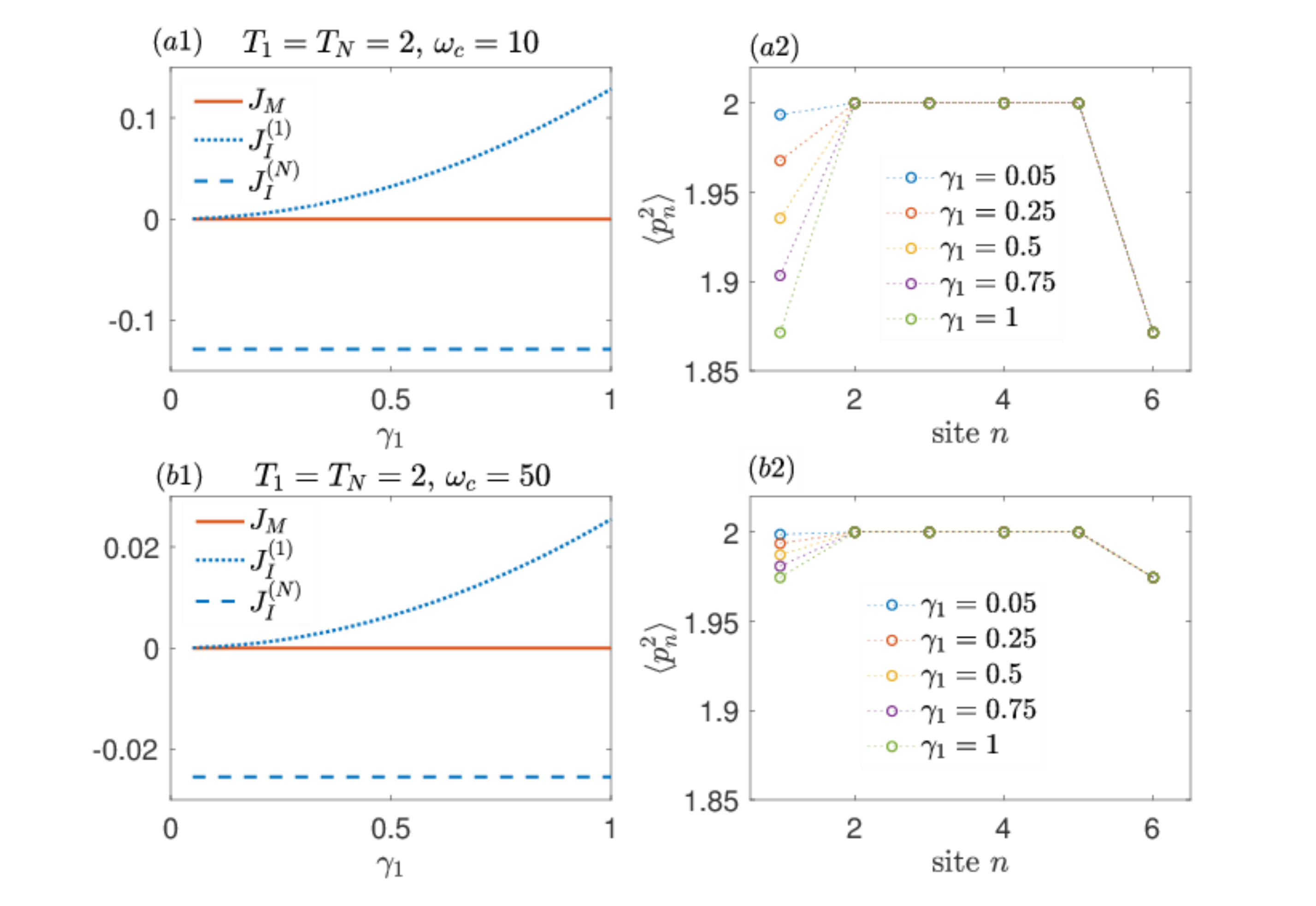}%{figeq2N2.eps}
\caption{Frequency domain equilibrium simulations with $T_1=T_N$. % and asymmetric junction, $\gamma_1\neq \gamma_N$.
(a1) Heat current and (a2) mean square momentum as a function of $\gamma_1$ for $\omega_c=10$
and (b1)-(b2) $\omega_c=50$. Here and below the integration step is $\delta \omega=5 \times 10^{-3}$.
The chain includes $N=6$ sites and we set $\gamma_N=1$ while varying $\gamma_1$.
Other parameters are $k=1$, $T_1=T_N=2$.
}
\label{Fig2}
 %\end{subfigure}
 \end{minipage}
%\end{figure}
%====================================
%  Figure 3
%\vspace{5mm}
%\begin{figure} [htbp]
\begin{minipage}[b]{0.45\textwidth}
%\begin{subfigure}[b]{0.4\textwidth}
\includegraphics[scale=0.3]{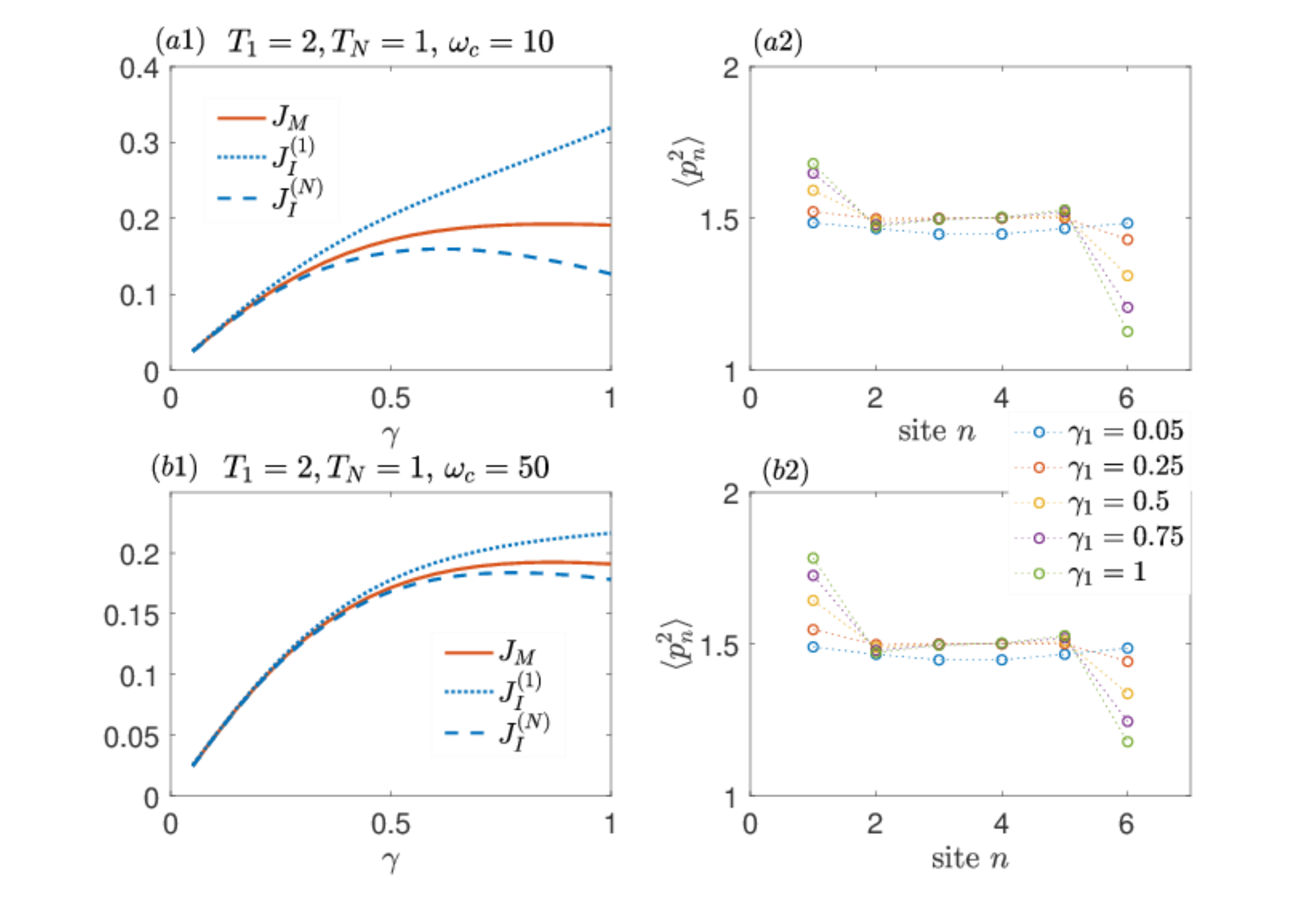}%{figneq1N2.eps}
\caption{Steady state heat current under a temperature bias $T_1=2$ and $T_N=1$, frequency domain simulations.
(a1) Heat current and (a2) mean-square momentum at $\omega_c=10$ and
(b1)-(b2) $\omega_c=50$ with the integration step $\delta \omega=5 \times 10^{-3}$.
The chain includes $N=6$ sites, $k=1$, and we vary the friction coefficient at the boundaries $\gamma=\gamma_{1,N}$ in a symmetric manner.
}
\label{Fig3}
\end{minipage}
% \end{subfigure}
\end{figure}
 %=================================================
 %===========================
%  Figure  4
\vspace{5mm}
\begin{figure} 
\includegraphics[scale=0.4]{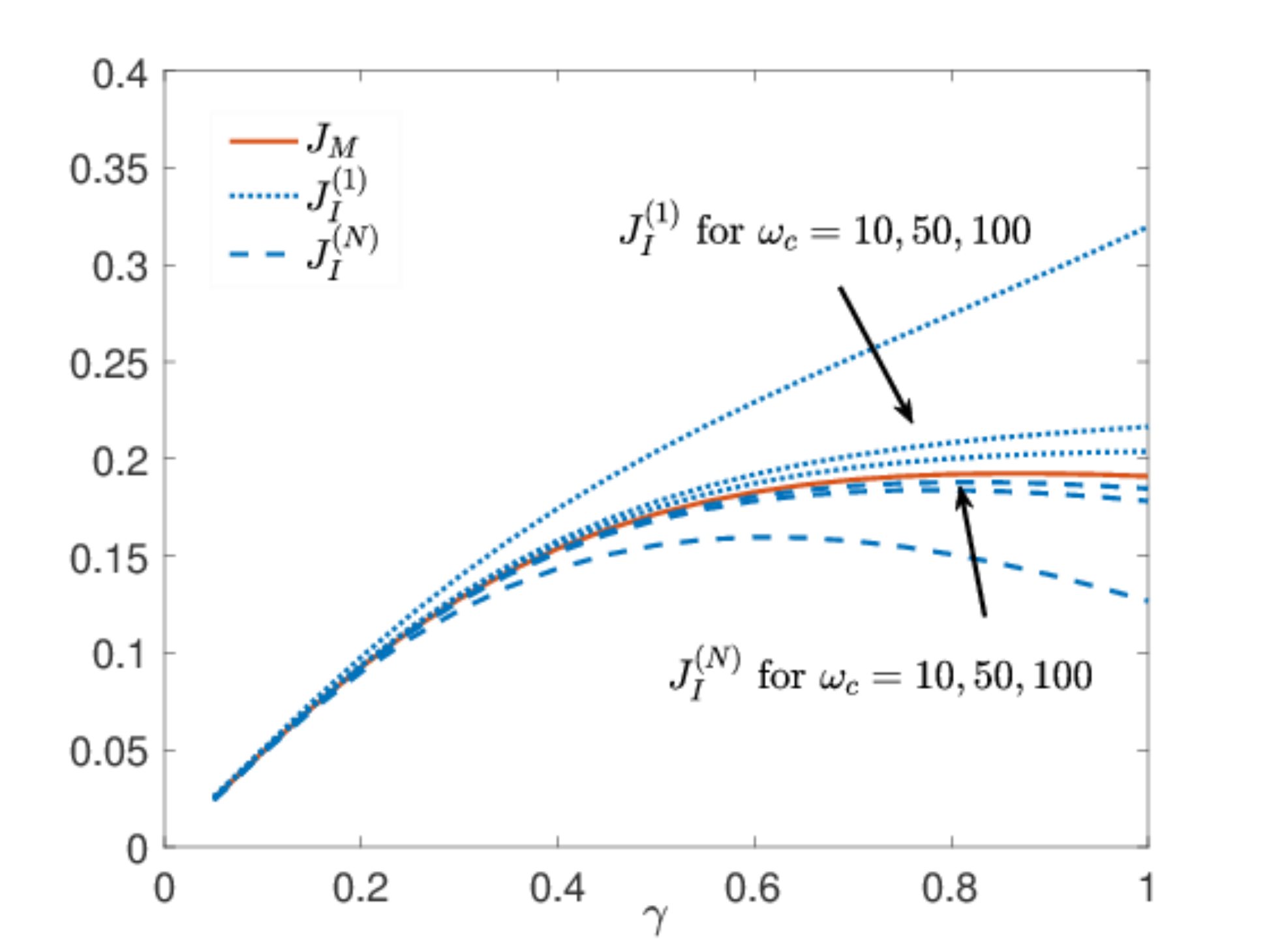}%{fig4N2.eps}
\caption{Convergence of the interface heat current upon increasing $\omega_c$, frequency domain simulations.
We use $\gamma=\gamma_{1,N}$, $T_1$ = 2, $T_N$ = 1, $N$ = 6, $k$ = 1.
The interface currents $J_I^{(1)}$ (dotted line) and $J_I^{(N)}$ (dashed) approach
the intramolecular definition (full) as we increase $\omega_c$.
The intramolecular current does not change with $\omega_c$.
}
\label{Fig4}
\end{figure}
%===========================
%  Figure 5
\vspace{5mm}
\begin{figure*}[hbtp] 
\includegraphics[scale=0.4]{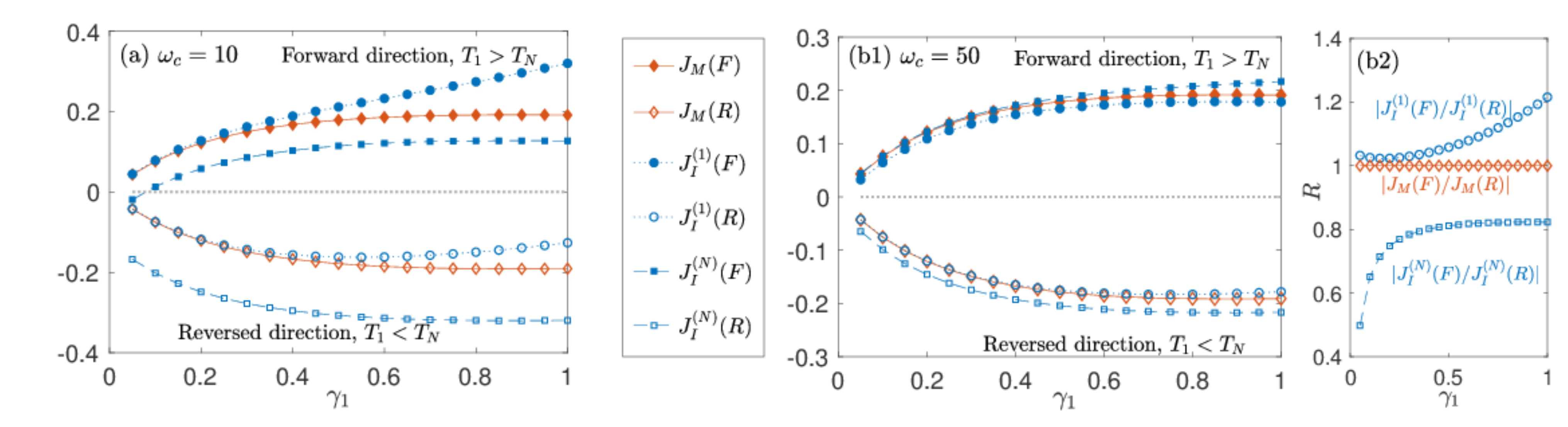}%{fig5N2.eps}
\caption{Erroneous thermal rectification effect for harmonic junctions based on the interface definition using frequency domain simulations.
We present the heat currents in the forward (F) and reversed (R) directions 
for $\omega_c$= 10 (a) and $\omega_c=$50 (b1). The magnitude of $J_I$ is different in the forward  and reversed  directions while 
$|J_M(F)|=|J_M(R)|$.
In panel (b2) we display the rectification ratio the for $\omega_c=50$ case. $R\equiv |J(F)/J(R)|$ is calculated by dividing currents in the forward direction $T_1>T_N$ (filled symbols) by currents in the reversed direction $T_1<T_N$ (empty symbols).
We use  $T_1=2$, $T_N=1$, $N=6$, $k$=1, $\gamma_N$=1.
}
\label{Fig5}
\end{figure*}
%==============================
%=========================
%====================
\subsection{Simulations}
 
We simulate heat transport with the intramolecular definition $J_M^{(n)}$ using Eq. (\ref{eq:JBHA}). It can be evaluated anywhere in the chain, and we use the chain's center (results were identical for any $n$). 
The interface definition at site 1, $J_I^{(1)}$ is calculated from  
Eq. (\ref{eq:JIH});  an analogous expression is used for the interface current at the other end, $J_I^{(N)}$.
In simulations we set $\pm\omega_c$ as the limits of integration.
 The integration interval $\delta \omega$ was taken small enough; we confirmed that it did not introduce an integration error.
 Regarding units used: For simplicity in simulations $k_B=1$, $\hbar=1$. Therefore, 
 $\omega_c$, $\gamma$, $T$, and $\sqrt{k/m}$ have the same units (energy). The simulated heat current has units of energy squared. If, for example we set  the energy unit to 20 meV, we get $T=1\to$ $k_BT=20$ meV, 
 $\gamma=0.1 \to$ $\gamma =0.5$ ps$^{-1}$, %(0.1*20*1e-3*1.6e-19/6.626e-34)*1e-12
  $\sqrt{k/m}=1 \to$ $\sqrt{k/m}=160$ cm$^{-1}$ %(20e-3*1.6e-19/6.626e-34)/3e10
 and the resulting current, $J=0.2 \to $ $\sim 20$ nW. % 0.2/(6.626e-34/2/pi)*1e-6*1.6e-19*1.6e-19*400

{\it Thermal equilibrium.}
We test the equilibrium scenario in Fig.~\ref{Fig2}. While  $J_M$ is  identically zero irrespective of the value of $\omega_c$, we find that $J_I$  shows a significant error, a nonzero current (compare magnitude to Fig.~\ref{Fig3}).
We further study the mean-square momentum, which should be equal to $k_BT$ at equilibrium based on energy equipartition.
The kinetic energy deviates from the equipartition value at the boundaries, and this deviation reduces as we increase $\omega_c$. However, it is notable that $J_M=0$ irrespective of the accuracy of energy equipartition, while the quality of $J_I$ relies on this property. 
Essentially, in the intramolecular definition every frequency component in the current is individually cancelled out between the two heat baths. Therefore,  even if the upper limit of integration is not high enough to  achieve equipartition, the current is identically zero at equilibrium.

While in Figure \ref{Fig2} we use $\gamma_1\neq \gamma_N$, we emphasize that structural asymmetry is not the source of the problem.
When we repeat this simulation with $\gamma_1=\gamma_N$, we get that $J_I^{(N)}$ becomes a mirror image of $J_I^{(1)}$, 
with both currents significantly deviating from the correct (zero current) value.

%=================================

{\it Nonequilibrium steady state.}
In Fig.~\ref{Fig3} we apply a temperature difference across the junction and compare the cases $\omega_c=10$ and $\omega_c=50$. 
While $J_M$  does not depend on the integration limits at this resolution, and it is well converged (see also Fig.~\ref{Fig4}), we find that $J_I^{(1)} $ and $J_I^{(N)}$ significantly deviate from the correct answer, showing a strong dependence on 
$\omega_c$. Furthermore, when we plot
the mean squared-momentum  (panels a2 and b2) we find that modifying $\omega_c$ slightly changes the value at the boundary atoms, 1 and $N$.
% again that all sites deviate from equipartition, though 
It is notable that  a 5\% modification in $\langle p_1^2\rangle$ (as we tune $\omega_c$ from 10 to 50) translates to about $50 $\% shift in the magnitude of the interface current, illustrating its strong sensitivity to the average energy at the boundaries.

We further demonstrate the slow convergence of $J_I$ with $\omega_c$ in Fig.~\ref{Fig4}. While $J_M$ converges  once  $\omega_c\approx10 k_BT_a$ (the full line for $J_M$ does not vary as we increase $\omega_c$),
  $J_I^{(1)}$ and $J_I^{(N)}$ still visibly deviate from the correct answer even
 for $\omega_c\approx100k_BT_a$; $T_a$ is the average temperature.

%==============================

{\it Erroneous thermal rectification.}
Harmonic junctions connected to two heat baths at different temperatures cannot rectify heat \cite{Rego98,Segal03,SB,PereiraRev}. That is, the magnitude of the heat current should be the same when interchanging the temperatures of the reservoirs, $T_1$ by $T_N$. 
%This is apparent from the bulk current expression (\ref{eq:JBHA}), given in the form of a Landauer equation. 
%
 Fig.~\ref{Fig5} shows that the rectification ratio is identically one when adopting the intramolecular definition, 
$  R\equiv |J_M (F )/J_M (R)|=1$. However, the interface current shows a (faulty) significant rectification, up to a factor 2 if the integration is not carried out to large enough $\omega_c$. In panel (b) we repeat the calculation for a larger $\omega_c$. While there is a marked improvement with the interface currents approaching the intramolecular definition, the rectification ratio is still substantial as we present in panel (b2).

%==============================
\subsection{Error Analysis}
What is the reason for the puzzling-faulty behavior of the interface current?
In short, the interface current can be converted to the Landauer formula only after using a sum rule, which is difficult to converge. In contrast, the intramolecular current reduces to the Landauer formula based on an identity for the integrands. 
Let us play with Eq. (\ref{eq:JIH}):
\bea
%&&\gamma_1\left(  \langle p_1^2\rangle  -k_BT_1 \right)
%\nonumber\\
&&\sum_m\frac{\gamma_1\gamma_m}{\pi}\int_{-\infty}^{\infty} d\omega\omega^2  \left|(G^r(\omega))_{1,m} \right|^2k_BT_m -\gamma_1k_BT_1
\nonumber\\
&&=
\sum_m\frac{\gamma_1\gamma_m}{\pi}\int_{-\infty}^{\infty} d\omega\omega^2  \left|(G^r(\omega))_{1,m}\right|^2 k_B\left(T_m - T_1\right)
\nonumber\\
&&+
\sum_m\frac{\gamma_1\gamma_m}{\pi}\int_{-\infty}^{\infty} d\omega\omega^2  \left|(G^r(\omega))_{1,m}\right|^2 k_B T_1-\gamma_1 k_BT_1.
\nonumber\\
&&=
 \sum_m\frac{\gamma_1\gamma_m}{\pi}\int_{-\infty}^{\infty} d\omega\omega^2  \left|(G^r(\omega))_{1,m} \right|^2k_B\left(T_m - T_1\right).
 \nonumber\\
 \label{eq:NN}
\eea
The last step relies on the normalization condition, Eq. (\ref{eq:N}).
With these simple manipulations, we received the multi-terminal Landauer formula for heat transport.
If only the first and last sites couple to thermostats, this last expression precisely corresponds to the two-terminal Landauer formula, $ J_{Land}= \frac{\gamma_1\gamma_N}{\pi}\int_{-\infty}^{\infty} d\omega\omega^2  \left|(G^r(\omega))_{1,N} \right|^2k_B\left(T_1 - T_N\right)$  \cite{Rego98,Segal03,ohmic}.

Obviously, the Landauer formula for heat currents does not suffer from pathologies such as a nonzero current at thermal equilibrium or a manifestation of a diode effect---given its explicit dependence on the temperature difference.
We have therefore just proved that
$J_I$ of Eq. (\ref{eq:JIH}) is mathematically equivalent to the good-old Landauer formula. Why then does Eq. (\ref{eq:JIH})
show computational flaws, e.g. a nonzero current at equilibrium?
The answer is that in transforming Eq. (\ref{eq:JIH}) to the Landauer form we relied on the normalization integral (\ref{eq:N}).
To achieve an accurate normalization, the limits of integration ($\omega_c$) need to be extended to very high frequencies. %For example,  while $J_M$ excellently converges for $\omega_c\approx 10k_BT$, 
Specifically, we need to use $\omega_c> 100k_BT$ to get
an error smaller than 10\%, that is $|J_I^{(1)} /J_I^{ (N )} | <1.1$.

From Eq. (\ref{eq:NN}) we also note that
since $\sum_m\frac{\gamma_1\gamma_m}{\pi}\int_{-\omega_c}^{\omega_c} d\omega\omega^2  \left|(G^r(\omega))_{1,m}\right|^2 <\gamma_1$, the interface current at site `1'  should exceed the correct value for $T_1>T_N$ when un-converged, $J_I^{(1)}>J_M$, and the other way for $J_I^{(N)}$.

Complementing this discussion, in Appendix B we prove that the intramolecular current (\ref{eq:JBHA}) reduces to a Landauer form with an explicit dependence on the temperature difference by utilizing a trivial identify for the integrands---that does not depend on the limits of integrations---thus does not suffer from numerical pathologies. 

%We now demonstrate this problem by presenting simulations based on numerical integration of Eqs. 
%Eq. (\ref{eq:JBHA}) and Eq. (\ref{eq:JIH}) for
%for $J_B^{(n)}$, $J_I^{(1) }$, along with the corresponding expression for $J_I^{(N)}$, while truncating the limits of integration at $\pm\omega_c$.

%The bulk definition offers two appealing features, of physical importance.
%First, at equilibrium the current is identically zero. 
%Second, it is obvious that in a two-terminal problem, switching the temperatures 
%$T_1$ by $T_N$ results in currents of identical magnitude (no diode effect). 

%Since the error collected in this numerical integration is different when calculated for different sites, i.e., $N_1\neq N_4$. As such, 
%averaging over sites does not resolve this issue.

%====================================
% Figure 6
\begin{figure*} 
\includegraphics[scale=0.4]{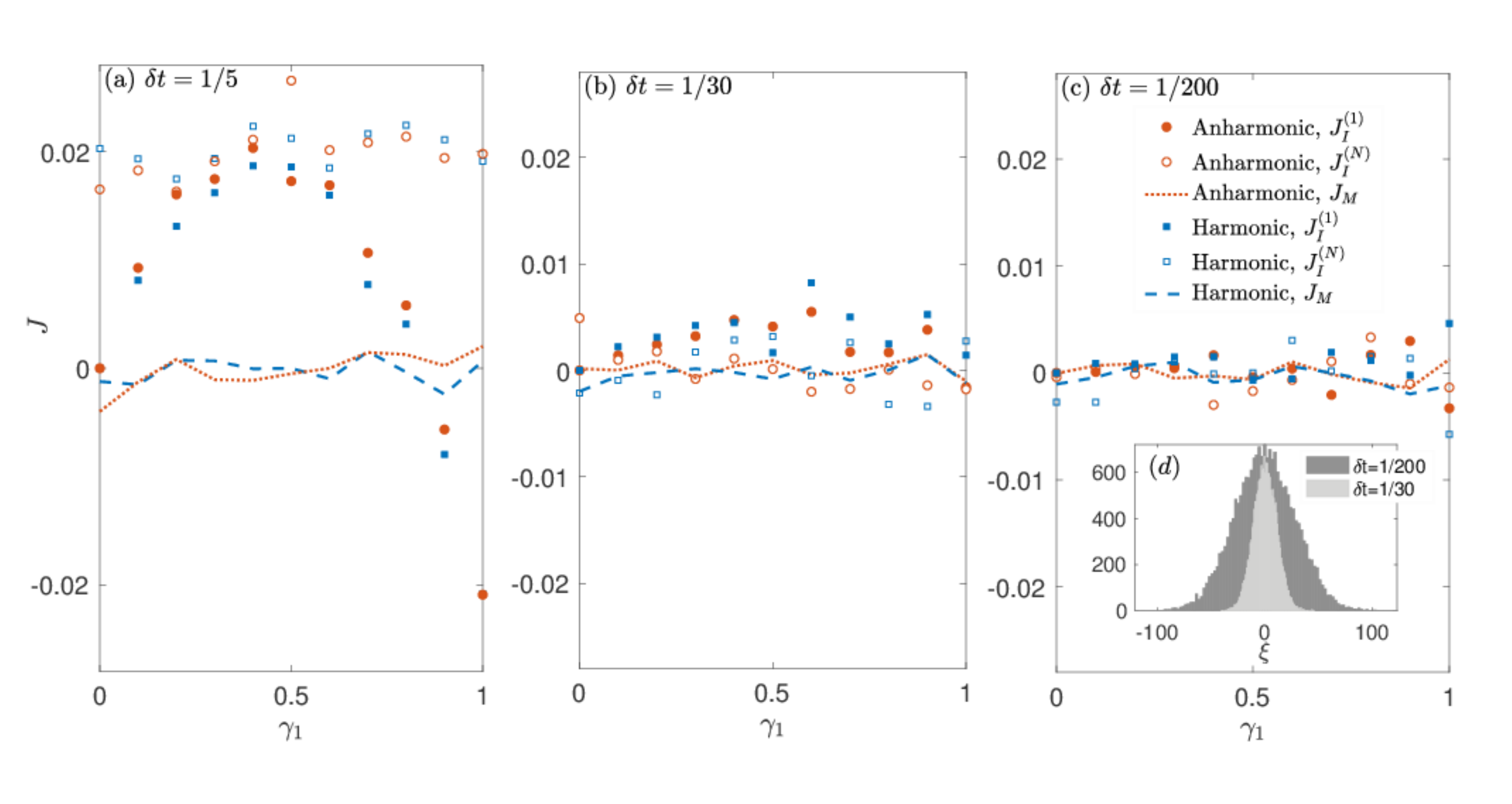} %{eq_fric_N1.eps}
\caption{Molecular dynamics simulations of an equilibrium setup with $T_1=T_N=1$. 
(a)-(c) Heat current as a function of $\gamma_1$ for harmonic and anharmonic chains.
The intramolecular current is close to zero even for a rough time step, while the interface currents show substantial errors, which 
can be reduced when taking a smaller time step.
%Note that the scale in panels (a) and (b) is the same. %; we zoom over the current of panel (b1) in panel (b2).
Parameters are $N=6$, $\gamma_N=1$, $k=1$, and the integrator is RK4.
Unless otherwise mentioned, here and in figures below the time interval is $\tau=90$  and $J_M$ is calculated between sites 3 and 4.
(d) Histograms of the gaussian random noise generated for two different time steps, showing the broadening of the histograms as the time step becomes shorter.
}
\label{FigMD1}
\end{figure*}

% Figure 7
\begin{figure*} 
\includegraphics[scale=0.4]{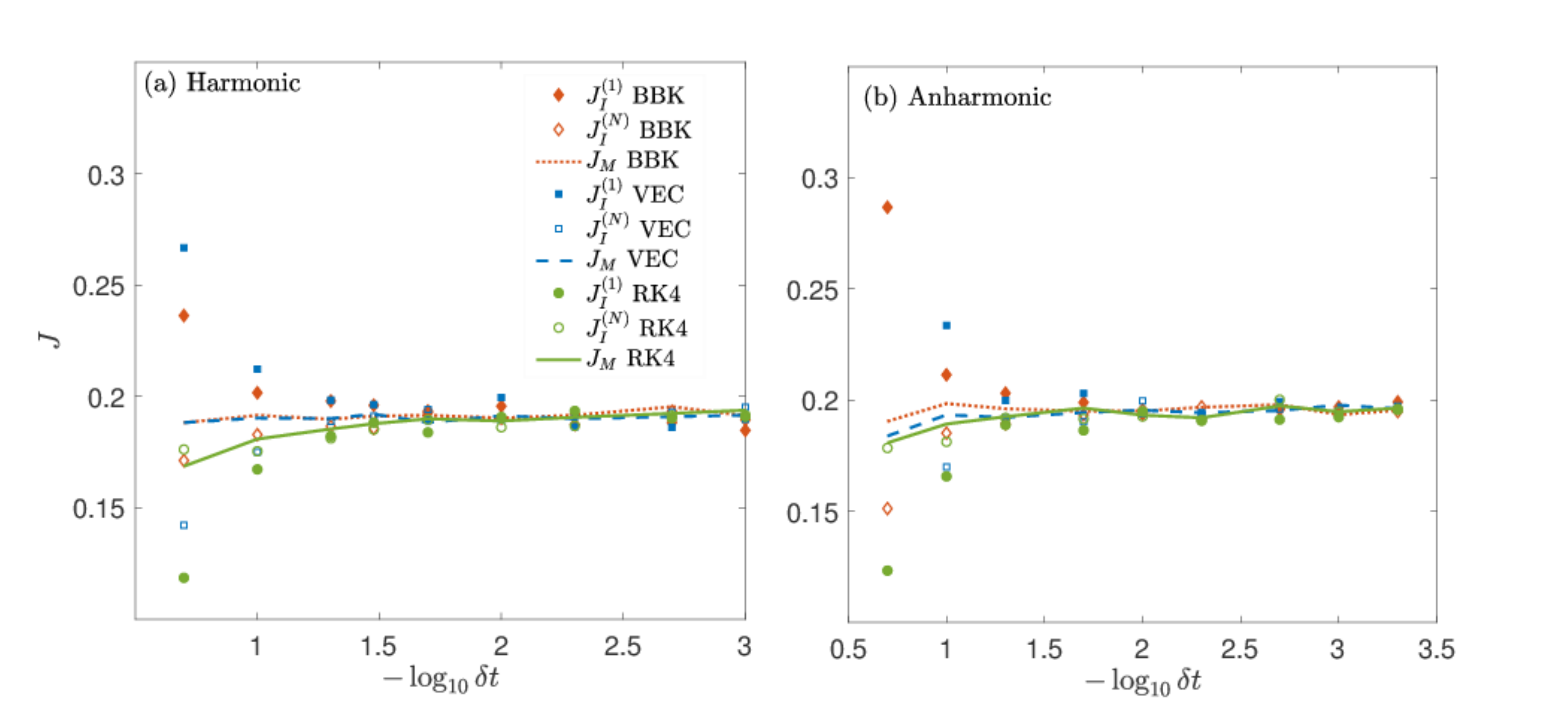} %{dt_errorN2.eps} % {dt_error_N1.eps}
\caption{Molecular dynamics simulations of heat currents in a nonequilibrium steady state for
(a) harmonic  and (b) anharmonic potentials.
%The interface currents converge only when $\delta t$ is much smaller than other timescales of the system. 
We use $T_1=2$, $T_N=1$, $N=6$, $k=1$, $\gamma_1=\gamma_N=1$ and we
test three integrators: RK4, BBK and VEC.   } 
\label{FigMD2}
\end{figure*}

% Figure 8
\begin{figure*} 
\includegraphics[scale=0.4]{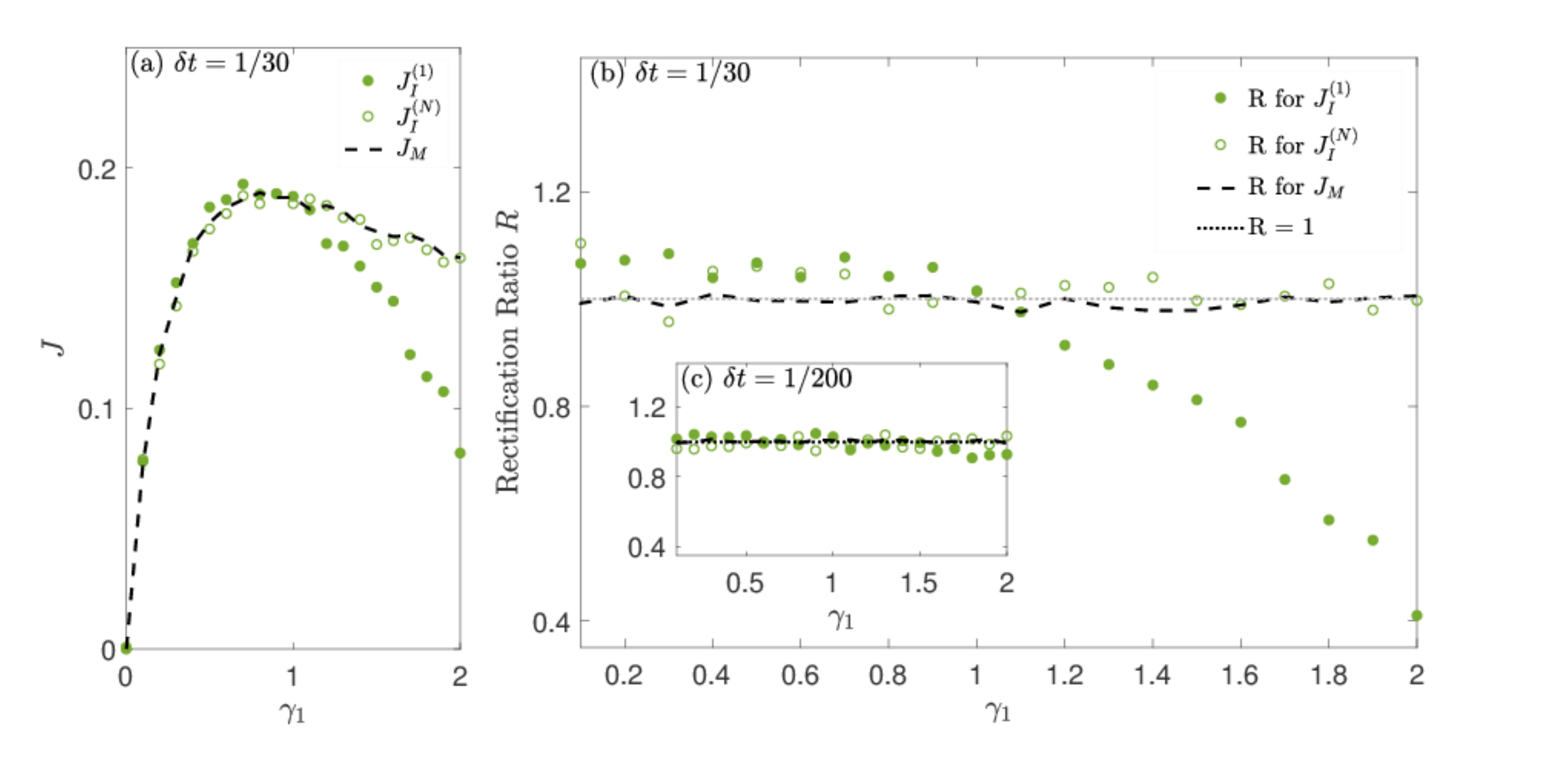} %{noneq_fric_N2.eps}
%\vspace{-20mm}
\caption{An erroneous thermal rectification effect in harmonic systems based on molecular dynamics simulations.
(a) Current and (b) erroneous rectification
using RK4 for $\delta t =1/30$. 
(c) Rectification disappears when using a finer time step of $\delta t =1/200$.
We use $N=6$, $\gamma_N=1$, $k=1$, $T_{H}=2$ and $T_{C}=1$.} 
\label{FigMDR}
\end{figure*}

%===============================

\section{Time domain: Molecular dynamics simulations}
\label{Sec-MD}

The exact Green's function method described above is limited to harmonic models.
While a Meir-Wingreen type expression can be derived for anharmonic systems \cite{Thoss}, calculations of the
Green's function involved can only be done perturbatively for the anharmonic contributions \cite{BijayGRev,BijayRev}.
In classical systems, molecular dynamics simulations are therefore  central for the study of phononic heat conduction.

In this Section we simulate phononic conduction in harmonic and anharmonic molecular junctions
using classical molecular dynamics simulations
 and testing the performance of the two (intramolecular, interface) definitions. 
We set the initial conditions, then time-evolve the equations of motion for long enough production time to get a trajectory corresponding to the steady state limit. 
We numerically integrate the Langevin equation (\ref{eq:EOM}) testing three different integrators, the fourth-order Runge-Kutta (RK4) method, the Br\"unger-Brooks-Karplus (BBK) integrator \cite{BBK}, and with the method developed by Vanden-Eijnden and Ciccotti \cite{Ciccotti}, which we refer to as the VEC integrator.

To compute heat currents, we wait for the system to reach a steady state, then at each time step (duration $\delta t$)  within a time interval $\tau$ we record the heat currents. To take an average over realizations of the noise we repeat this process over an ensemble of size $N_\zeta$, and we average the $\frac{\tau} {\delta t} \times N_\zeta$  values of currents. We consider both harmonic and anharmonic potentials. In the anharmonic case we use $V(x_1,\dots, x_n) = \frac 14 \sum_{n=2}^{N} k (x_n - x_{n-1} - a)^4$ for the interparticle potential energy, and the same form for the contact coupling $V_B$.
Unless otherwise specified, we set $k=1$ in the potential energy, use $\gamma=0.1-2$,  and average over an ensemble size $N_\zeta = 4000$ and a time interval $\tau = 90$, while using a simulation time step in the range $\delta t=10^{-3}-0.5$.

Our main finding in this section is that, similarly to observations in the frequency domain, the intramolecular heat current is advantageous over the interface currents. The intramolecular current converges faster when decreasing $\delta t$ than the interface current.
% which shows significant deviations from the correct value until $\delta t$ is significantly smaller than $\gamma^{-1}$, $(k_BT)^{-1}$, and $k^{-1/2}$. 
While the intramolecular definition correctly shows a vanishing current at equilibrium and the absence of thermal rectification effect for harmonic junctions, for corresponding parameters the interface currents display finite current at equilibrium, and a rectification effect.
Overall, we argue that the impact of the propagation time step ($\delta t$) error in molecular dynamics simulations is analogous to the role of the frequency cutoff $\omega_c$ in the frequency domain. 

%Simulations presented in the main text, Figs. \ref{FigMD1}-\ref{FigMDhist} were generated with the RK4 integrator. Appendix C includes simulations utilizing the BBK and the VEC integrators, but  results based on these integrators were similar: In all cases we found that the intramolecular current shows a faster convergence over the interface currents. 

\subsection{Averaged heat current}

{\it Thermal equilibrium.}
In Figure \ref{FigMD1} we study the equilibrium scenario with $T_1=T_N$ using the RK4 method.
We find that the intramolecular current is very close to zero even for a rough time step.
In contrast, for the same time step discretization, the interface expressions at the left and right sides display significant erroneous heat currents. 

We make the following observations:
(i) The erroneous interface current is suppressed when reducing the time step from $\delta t=0.2$ to $\delta t=0.03$.
This improvement is not due to the additional averaging associated with increasing the trajectory time, $\tau/\delta t$ for a shorter time step. Indeed we verified that keeping $\delta t$ fixed while increasing $\tau$ does not suppress the flawed current $J_I$.  %XXX
(ii) The interface definition inadequately performs  (for large $\delta t$) in both harmonic and anharmonic systems. 
(iii) In some cases, a symmetrization of the interface current may assist to reduce its error, yet it does not cancel the error exactly.
%, and the averaged interface current $(J_I^{(N)}-J_I^{(1)})/2$ still has an error greater than the intramolecular current.
%(iv) The numerical error in $J_I$ seems to grow with $\gamma_1$, in correspondence with Fig.~\ref{Fig2}.
%
(iv) As we continue and reduce the time step from $\delta t=1/30$ to $\delta t =1/200$, we do not observe additional improvements in the 
behavior of the heat current, demonstrating that this residual error is associated with the ensemble average.
We note that as we reduce the  time step, more intense random forces are taken into account,  see the histogram of the gaussian noise in panel (d) of Fig.~\ref{FigMD1}.  
We performed additional simulations for other chain lengths, $N=4-10$ and confirmed that our observations were unchanged. Generally, it was more difficult to converge the interface current with increasing chain length. % DDD Repeat

{\it Steady state heat current.}
In Fig.~\ref{FigMD2} we study the currents in harmonic and anharmonic cases under a temperature bias. First we confirm that  the harmonic intramolecular current agrees with calculations done in the frequency domain, see Fig.~\ref{Fig3}. 
We find that the error of the interface currents  drops when reducing the time step. 
Note that  since the coefficient $k$ in the potential energy has a different physical dimension for harmonic and anharmonic junctions, 
we cannot meaningfully compare these results. 
 A relevant comparison could be made by studying a model with harmonic plus anharmonic terms,
slowly turning on the anharmonicity. 

We test three different integrators and conclude that independently of the integrator, the intramolecular current definition is advantageous over the interface current, as the latter requires more computational effort (shorter time step) to converge to the correct result. 
It is intriguing to note that in the BBK and VEC methods  $J_I^{(1)}>J_M$, while the opposite holds for RK4. To explain these trends one needs to carefully study errors associated with the different integrators, and how these errors impact energy equipartition. 

{\it Thermal rectification.}
We test the (faulty) development of the thermal rectification effect in harmonic junctions in Fig.~\ref{FigMDR}. 
First, in panel (a) we display the  heat current as a function of $\gamma_1$. 
 These simulations were performed in the ``forward" direction, with $T_1>T_N$. 
Next, in panels (b)-(c), we analyze the rectification ratio $R$. While the intramolecular current 
shows no rectification, the interface current $J_I^{(1)}$ displays the effect, and it can be substantial for large asymmetry.
We repeated these simulations while fixing $\gamma_1$ and varying $\gamma_N$. In this case, large rectification shows at the other contact current, $J_I^{(N)}$.
In Ref. \cite{Muga} it was argued, based on molecular dynamics simulations, that a linearized (harmonic) model could support the thermal diode effect. We point that this observation may result from the pathologies of the interface current (employed in that study), when improperly  converged. 
 %\textcolor{red}{What about the anharmonic case?}
%==============================
% Figure 9
\begin{figure*} 
\includegraphics[scale=0.4]{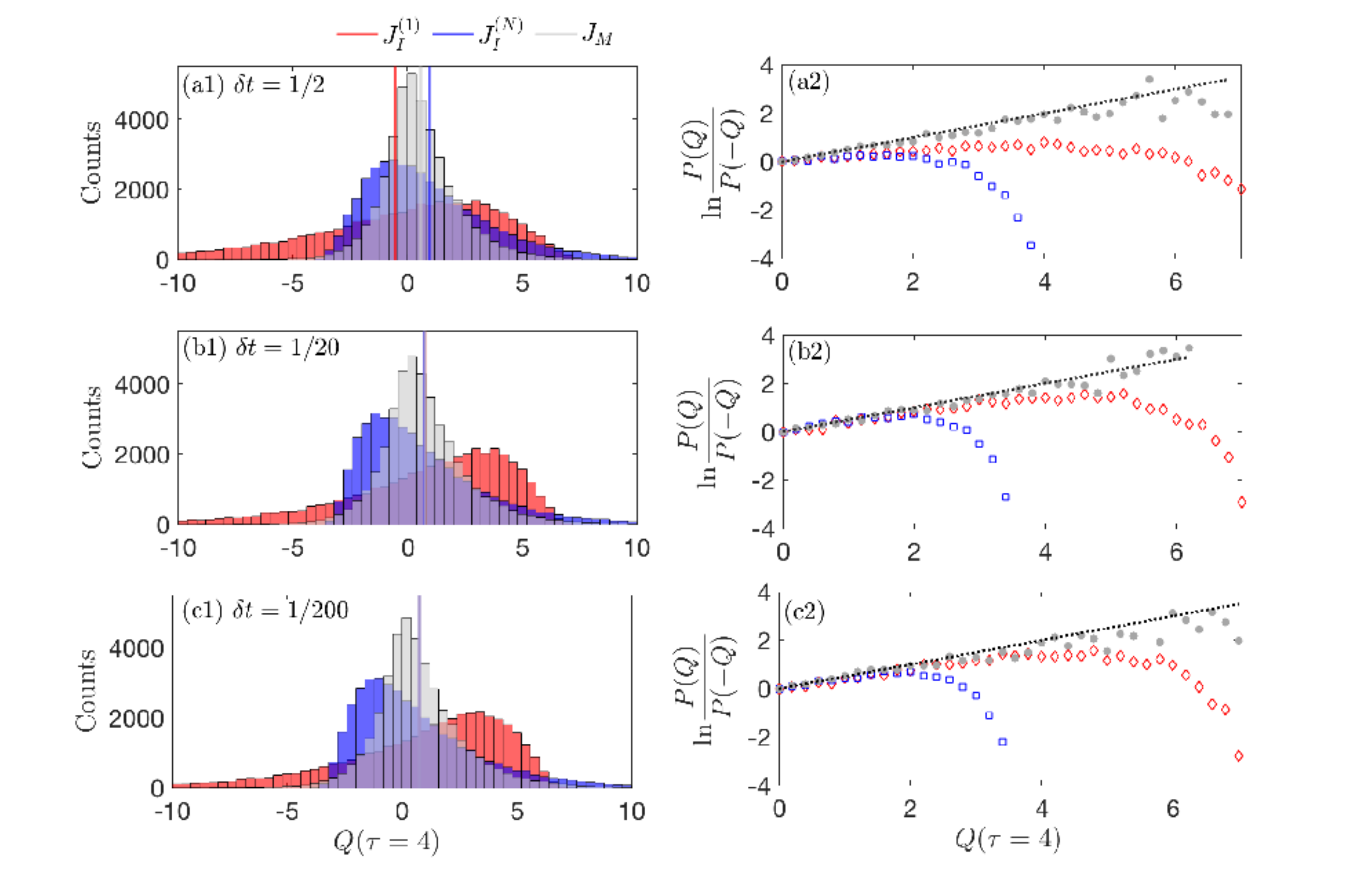}%{big_tau_RK_histN.eps} % {histN2.eps} %{new_BBK_hist_N1.eps} %{histN1.eps} %{hist-dt-20-200-2000-N1.eps}
\caption{(a1)-(c1) Probability distributions of observed integrated heat currents $P(Q)$ within a duration $\tau$ using the RK4 integrator. 
%The vertical lines mark the mean of the distributions. 
Reducing $\delta t$ does not much alter the shape of the distribution, but it causes the means (vertical lines) to converge.
The interactions in the chain are harmonic with and we adopt $k=1$, $\gamma_{1,N}=1$, $N=4$, $T_1=2$ and $T_N=1$. %N=4?DDD
In panels (a2)-(c2) we test the fluctuation symmetry. It approaches the correct slope (dotted) for the intramolecular current,
$\Delta\beta=1/T_C-1/T_H=1/2$, while for the interface currents deviations show up even as the averaged current converges to the correct value.
}
\label{FigMDhist}
\end{figure*}

%===============================

\subsection{Heat exchange fluctuation relation}
Beyond the averaged current, we are interested in the full probably distribution function of heat exchange.
This would allow us to confirm that the simulation protocol is correct---satisfying the steady state exchange fluctuation symmetry,  $ P(Q)/P(-Q) = e^{\Delta \beta Q}$ \cite{FR}. Here, 
$Q$ the integrated heat current within a certain duration, 
 $P(Q)$ is the probability distribution function of heat exchange $Q$, and
$\Delta \beta=1/T_C-1/T_H$, % We need to use here Th and TC, not 1 and N, because the FR does not make sense with 1 and N 
$T_C$ ($T_H$) is the temperature of the cold (hot) thermal bath.
 
To produce a histogram of the heat exchange, we 
follow many trajectories of duration $\tau$ and calculate the integrated currents
 $Q(\tau)\equiv\int_{\tau_0}^{\tau_0+\tau} J(t)dt$. The time $\tau_0$ is taken long enough so as initial conditions become irrelevant.
We calculate the integrated currents using the intramolecular and interface definitions, and construct the three histograms.

First, we point out that to calculate the integrated interface current, $Q_I(\tau)$, we  cannot use  Eq. (\ref{eq:JI}),
 since this expression already relies on performing the average  $\langle \xi_n(t)p_n(t)\rangle$. Indeed, based on 
the defintion $J_I^{(1)}=-\gamma_1 \left(\langle p_1(t)^2\rangle/m_1 - k_BT_1 \right)$,
the stochastic current $J_I^{(1)}$ cannot be larger than $\gamma_1k_BT_1$. Similarly, the stochastic current at the other end, $J_I^{(N)}$ cannot take values below $-\gamma_Nk_BT_N$. To properly generate $P(Q_I)$ we therefore resort to the original definition of the interface current, e.g. 
$J_I^{(N)}= \frac{\gamma_N}{m_N} \left\langle p_N(t)^2\right\rangle - \frac{1}{m_N}\langle \xi_N(t)p_N(t)\rangle$.
% which does not enforce that the Langevin forces are realizations of a gaussian white noise.  
%
However, in simulations we encounter an additional challenge in calculating $\langle \xi_n(t)p_n\rangle$, which we now explain.
Recall that in sophisticated  integrators one generates the value at $t+\delta t$ based on intermediate calculations, between $t$ and $t+\delta t$.
 For example, in the BBK scheme we first calculate the momentum at the half interval, $t+\delta t/2$ (the update is more complex in the RK4 method).
As such, it becomes unclear whether in the average  $\langle \xi_n(t)p_n\rangle$ one should take the momentum at the midpoint
$t+\delta t/2$, or after a full time step $t+\delta t$. From simulations we noted that this average delicately depends on this choice; in both RK4 and the BBK methods, we found that to produce the correct average, $\langle \xi_n(t)p_n\rangle=\gamma m_n k_BT_n$,
the momentum had to be selected at the midpoint, $t+\delta t/2$.
% DDD recheck

In Fig.~\ref{FigMDhist}, we present histograms of heat exchange using three different time steps. We further test the fluctuation symmetry 
 at the right panels.
We find that the intramolecular current obeys the fluctuation symmetry, $P(Q)/P(-Q) = e^{\Delta \beta Q}$ even with a rough time step.
% That is, the probability that within $\tau$ 
%$Q(\tau)$ will be extracted from the cold bath is exponentially suppressed relative to the opposite flow process.
 In contrast, the interface currents   violate this symmetry, {\it even when the averaged currents are converged once the time step is short enough in panels  (b) and (c)}.
We conclude that in molecular dynamics simulations, the interface definition should not be used to study the full counting statistics:
 (i) Its construction is ill-defined (as mentioned above, we used momentum at the half step to enforce the correct average).
 (ii) The shape of the distribution is incorrect even when the mean converges to the correct steady state value.
 % Turn (a) into 1/10? DDD

\subsection{Comparison between integrators}
Results presented in Figs.~\ref{FigMD1}, \ref{FigMDR} and \ref{FigMDhist} were obtained using the RK4 integrator. 
In Appendix C we include additional demonstrations using the BBK integrator \cite{BBK}. % integrators \cite{Ciccotti}.
The three integrators show that the intramolecular current definition is advantageous over the interface current: the latter requires more computational effort (shorter time step) to converge to the correct result, and its distribution violates the fluctuation symmetry.
%Of course, a high-order integrator such as the VEC exemplifies superior convergence over BBK and RK4. However, the convergence problem of the interface current is still presented.
%
We recall that the two definitions of the current involve different correlation functions (position-velocity in the intramolecular
and velocity-squared at the boundaries).
While previous studies compared the performance of different integrators against each other, see e.g. Ref. \cite{Wei}, our analysis here concerns the convergence of different correlation functions with a given integrator. 
It is  interesting to note that the RK4, while not specifically designed to propagate a stochastic differential equation, converges the intramolecular current very well. We further note that the Runge-Kutta integrator is commonly used in fundamental physics studies of anomalous  heat transport in one-dimensional systems; some prominent studies include \cite{casati02,casati04,bambi06}. % add 
In contrast, the BBK and the VEC integrators are popular in molecular dynamics simulations of biomolecules and materials.

%==============================
\section{Summary}
\label{Sec-Summ}

We focused on the problem of phononic heat transport in nanojunctions and asked an elementary question: Which definition for the heat current best performs in simulations?
We argued that while the interface definition is advantageous in formal derivations and in the quantum domain, in classical simulations one should employ the intramolecular definition as it shows a superior convergence.

Considering harmonic junctions, we simulated heat current in the frequency and time domains. 
Frequency-domain calculations were done
in the language of Green's functions. We showed that the interface definition was poorly converged when truncating high frequency modes, since it relied on a sum rule to converge. In contrast, by construction, the intramolecular current showed favorable properties: The heat current was identically zero at equilibrium and a diode effect was disallowed.

In the time domain, we performed corresponding molecular dynamics simulations and demonstrated that the interface definition required a finer time step to converge, compared to the intramolecular expression.  
The deficiencies of the interface current manifested themeselves in an unphysical behavior of the heat current at equilibrium,
with a difficulty to achieve equipartition of energy close to the interface,
 and in the development of a diode effect for harmonic junctions.
These issues were observed with different integrators, RK4, BBK and the VEC. 
Insights on the poor performance of the interface currents could be achieved based on the upside/downside statistical analysis suggested in Refs \cite{NitzanUD1,NitzanUD2}. % DDD
While so far much effort had been placed on comparing the performance of different integrators, here 
we emphasize that  using different working definitions for an observable may result in different convergence behavior.

%Since our focus here had not been on comparing integrators, we did not perform a  comparative error analysis of the three integration schemes.

%Future work will be focused on the analysis of heat current in complex networks with beyond nearest-neighbors connectivities.
%The interface definition is prone to numerical inaccuracies, and thus to incorrect physical predictions.
Given significant advancements in experimental studies of vibrational heat transport across single molecules,  and similarly, progress in emulating transport problems with engineered chains such as trapped ions structures \cite{Haffner,Tamura},  it becomes increasingly important to perform accurate numerical simulations so as to bring useful predictions and gather fundamental understandings of mechanisms. Furthermore, to probe noise-precision trade off relations in thermal machines, we need to simulate both the current and its fluctuations
in a nonequilibrium steady state \cite{TUR-udo,TUR-dechant,TURH}. 
As we showed here,  
the intramolecular definition had correctly captured the fluctuations of heat exchange by delivering the steady-state fluctuation symmetry,  therefore it acted as a benchmark.
In contrast, the interface definition of heat exchange showed violations of the fluctuation symmetry. Even when the averaged current converged to the correct result,  fluctuations were not correctly described by the interface expression. 
%As such, extra care must be taken in relying on molecular dynamics simulations when studying heat current fluctuations. 

 %Future work will be focused on studying the effect of long range interaction on thermal transport problems.

In principle, one could always reduce the time step in molecular dynamics simulations and simply verify convergence, irrespective of the definition employed. Our message is that: (i) Before convergence is reached, the interface definition leads
to incorrect physical predictions, unlike the intramolecular current. (ii) The interface definition cannot properly generate the full counting statistics of heat exchange. 
%
%In summary, while the interface definition of heat exchange is beneficial in formal-analytical studies of heat exchange, in molecular dynamics studies  one should employ the intramolecular definition to accurately capture the heat current and its moments.
It would be interesting to go beyond a delta-impulsive noise process and perform similar analysis for non-Markovian thermostats \cite{Segal08,Yun}.

%==============================
\begin{acknowledgments}
DS acknowledges the NSERC discovery grant and the Canada Chair Program.
BKA gratefully acknowledges the start-up funding from IISER Pune and 
the hospitality of the Department of Chemistry at the University of Toronto.
The authors acknowledge illuminating discussions with Junjie Liu.
\end{acknowledgments}

\vspace{5mm}

{\bf Data Availability.}
The data that support the findings of this study are available from the corresponding author upon reasonable request.
%===============================================================

%================================
% Appendix A
\renewcommand{\theequation}{A\arabic{equation}}
\setcounter{equation}{0}  

\section*{Appendix A: The normalization condition as a sum rule}
In this Appendix we show that the normalization condition  Eq.~(\ref{eq:N}) is in fact a strict sum rule condition for harmonic oscillator systems given by the Hamiltonian Eq.~ (\ref{Hamiltonian}), with the beads further coupled to independent heat baths. 
This is true for quantum and classical systems.
For generality, we present the argument in a quantum mechanical description, treating position and momentum as 
operators.
We first write down the standard definitions for the retarded and advanced Green's functions, % hbar DDD
\begin{eqnarray}
G_{ij}^r(t) & =& - \frac{i}{\hbar} \theta(t) \big \langle \big[x_{i}(t),x_{j}(0)\big] \big \rangle,  \nonumber \\
G_{ij}^a(t) &=&  \frac{i}{\hbar} \theta(t) \big \langle \big[x_{i}(t),x_{j}(0)\big] \big \rangle,
\label{eq:Gra}
\end{eqnarray}
where the operators are written in the Heisenberg picture. We further assume that the Green's functions are time-translational invariant and therefore depend only on one time argument. 
Upon taking the first derivative with respect to time for the Green's functions, one can simply receive the following relation
\begin{eqnarray}
\dot{G}_{ij}^r(t) - \dot{G}_{ij}^a(t) =   - \frac{i}{\hbar} \big \langle \big[p_{i}(t),x_{j}(0)\big] \big \rangle,
\end{eqnarray}
which for time $t=0$, leads to an interesting observation 
\begin{eqnarray}
\dot{G}_{ij}^r(t\!=\!0) \!-\! \dot{G}_{ij}^a(t\!=\!0) \!=\! \!  - \frac{i}{\hbar} \big \langle \big[p_{i}(0),x_{j}(0)\big] \big \rangle \!=\! -\delta_{ij}.
\nonumber\\
\end{eqnarray}
In matrix notation, this relation implies,
\begin{equation}
\dot{{\bf G}}^r(t\!=\!0) \!-\! \dot{{\bf G}}^a(t\!=\!0) ={\bf I}.
\end{equation}
In frequency domain, after performing the Fourier transformation, the above relation translates to 
\begin{equation}
\int_{-\infty}^{\infty} \frac{d\omega}{2\pi}\,  \omega \, {\bf A} (\omega) = {\bf I},
\label{spectral}
\end{equation} 
where ${\bf A}(\omega) = i \big[ {\bf G}^r(\omega) - {\bf G}^a(\omega)\big]$ is known as the spectral matrix. Note that this particular sum-rule is valid for arbitrary oscillator system. For harmonic junctions coupled to independent baths, one can further get a closed expression for the spectral matrix, given by
\begin{equation}
{\bf A}(\omega)= 2 \, \omega\,  {\bf G}^r(\omega) {\bf \Gamma}(\omega) {\bf G}^a(\omega),
\end{equation}
where ${\bf \Gamma}(\omega)$ is a $N \times N$ diagonal matrix with entries $\gamma_{n}$. The diagonal components  of Eq.~(\ref{spectral}) along with the above relation gives the sum rule in Eq.~(\ref{eq:N}).  
%=======================================
% Appendix B
\renewcommand{\theequation}{B\arabic{equation}}
\setcounter{equation}{0}  % reset counter
\section*{Appendix B: Properties of the intramolecular definition for harmonic systems}

\subsection{Derivation of the intramolecular current}
The intramolecular heat current in harmonic chains, evaluated between sites $n$ and $n+1$ is given by  \cite{Roy06,DharRev,SegalR}
\bea
&&J_M^{(n+1)}=
\nonumber\\
&&-k_{n,n+1}\sum_m \gamma_m k_BT_m
 \int_{-\infty}^{\infty} d\omega \frac{\omega}{\pi} {\rm Im}\left[
{\rm G}^r_{n,m} (\omega)  {\rm G}^{a}_{m,n+1}(\omega)  
\right].
\nonumber\\
\label{eq:JBHAA}
\eea
Here, $ {\rm G}^{a}=({\rm G}^{r})^{\dagger}$ thus $[{\rm G}^{r}_{n+1,m}(\omega)]^*= {\rm G}^a_{m,n+1}(\omega)$.
First, we derive Eq. (\ref{eq:JBHAA}) from the symmetrized Eq. (\ref{eq:JB}),
\bea
&&J_M^{(n+1)}= \frac{k_{n,n+1}}{2}\left(  \langle x_n \dot x_{n+1} \rangle  +  \langle  \dot x_{n+1} x_n\rangle \right)
\nonumber\\
&&=k_{n,n+1} {\rm Re}[\langle x_n \dot x_{n+1} \rangle].
\label{eq:eq1}
\eea
We define the greater Green's function as 
\bea
G^>_{l,m}(t,t') = \frac{-i}{\hbar} \langle x_l(t)x_m(t')\rangle,
\eea
and as such,
\bea
 \langle x_l(t)\dot x_{l+1}(t)\rangle = i\hbar \frac{\partial }{\partial t'} G^>_{l,l+1}(t,t')|_{t'\to t}.
\eea
% 
%DDD
So far,  averages correspond to expectation values over the initial state (since we work in the Heisenberg representation).
We now perform a time average, so as to reach the steady state (ss) limit, %DDD should we use a different notation?!
\bea
 \langle x_l\dot x_{l+1}\rangle_{ss} = i\hbar \int_{-\infty}^{\infty}\frac{d\omega}{2\pi} (i\omega )G^>_{l,l+1}(\omega).
\eea
Substituting this relation into Eq. (\ref{eq:eq1}) we get 
\bea
J_M^{(n+1)}= -\hbar k_{n,n+1} \int_{-\infty}^{\infty}\frac{d\omega}{2\pi} \omega  {\rm Re}[G^>_{n,n+1}(\omega)].
\label{eq:eq3}
\eea
For harmonic systems in the ohmic bath limit and at high temperature \cite{DharRev}
\bea
G^>_{n,n+1}(\omega) = \sum_m (-i) \frac{k_BT_m}{\hbar \omega} 2 \omega \gamma_m G_{n,m}^r(\omega) G_{m,n+1}^a(\omega).
\nonumber\\
\eea
We substitute this into Eq. (\ref{eq:eq3}) and retrieve  the intramolecular current (\ref{eq:JBHAA}).

%========================
% Figure 10 C1
\begin{figure*} 
\includegraphics[scale=0.45]{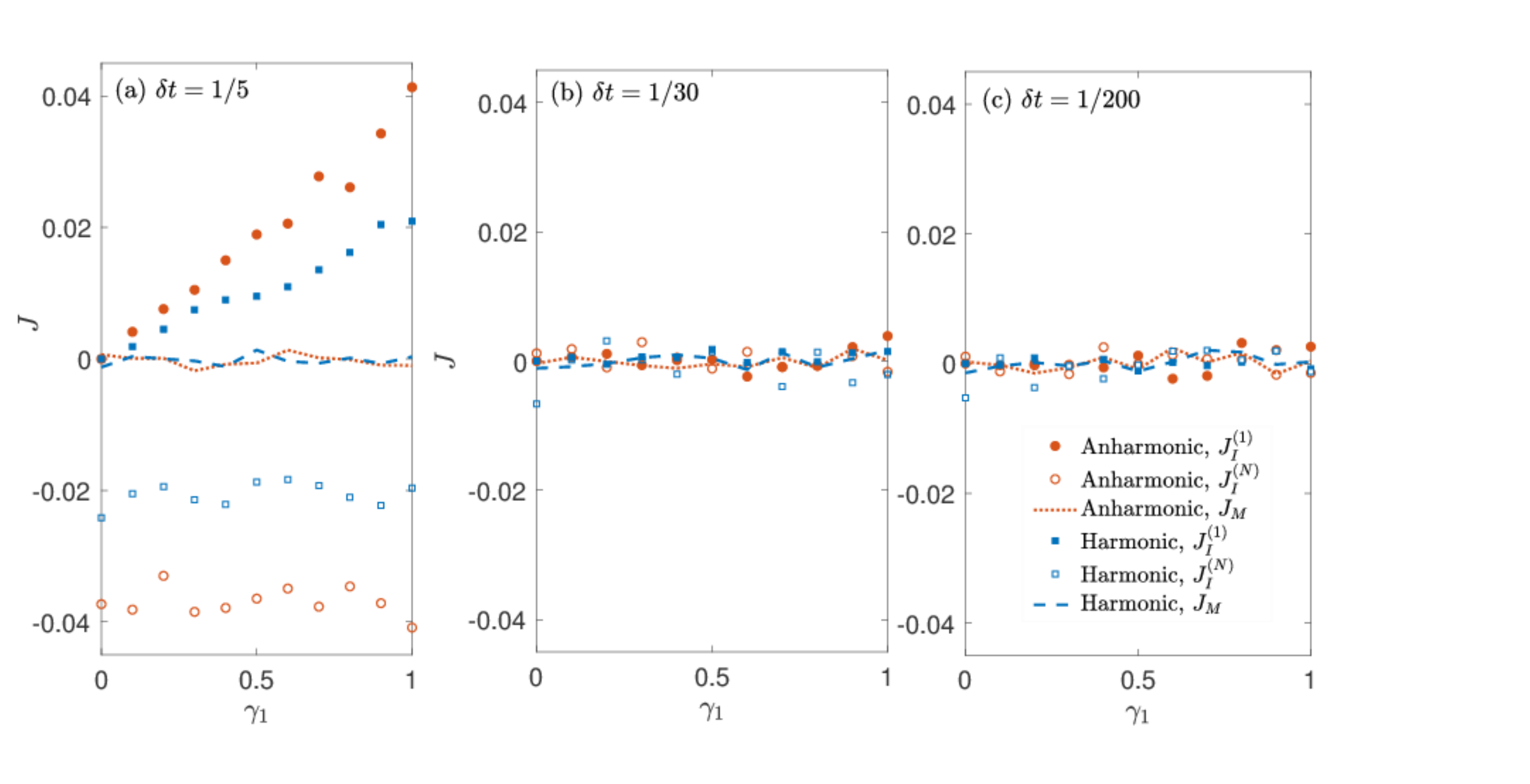}%{BBK_eq_fric_N2.eps}
\caption{Molecular dynamics simulations with the BBK scheme of an equilibrium setup with $T_1=T_N=1$. 
(a)-(c) Residual heat current  as a function of $\gamma_1$ for harmonic and anharmonic chains
with different time steps.
Parameters are $N=6$, $\gamma_N=1$, $k=1$.
}
\label{FigC1}
\end{figure*}
%
%=======================
\subsection{Proof that the intramolecular current is proportional to the temperature difference}

We note that at equilibrium, Eq. (\ref{eq:JBHAA}) reduces to
\bea
&&\sum_m \gamma_m 
 \int_{-\infty}^{\infty} d\omega \frac{\omega}{\pi} {\rm Im}\left[
{\rm G}^r_{n,m}(\omega)   {\rm G}^{a}_{m,n+1}(\omega)  
\right] = 0
\nonumber\\
\label{eq:N2}
\eea
We will discuss this identity in more details starting in Eq. (\ref{eq:N3}).

We now show that the intramolecular current is proportional to the temperature difference $(T_1-T_N)$, thus (i) it is identically zero at equilibrium,
and (ii) it cannot produce a diode effect.
Since in our system only $\gamma_1$ and $\gamma_N$ are nonzero, Eq. (\ref{eq:JBHAA}) becomes
\bea
&&J_M^{(n+1)}=
\nonumber\\
&&- \gamma_1 k_BT_1k_{n,n+1}
 \int_{-\infty}^{\infty} d\omega \frac{\omega}{\pi} {\rm Im}\left[
{\rm G}^r_{n,1}(\omega)   {\rm G}^{a}_{1,n+1} (\omega) 
\right]
\nonumber\\
&&-
\gamma_N k_BT_Nk_{n,n+1}
 \int_{-\infty}^{\infty} d\omega \frac{\omega}{\pi} {\rm Im}\left[
{\rm G}^r_{n,N} (\omega)  {\rm G}^{a}_{N,n+1}(\omega)
\right],
\nonumber\\
\eea
which can be written as 
\bea
&&J_M^{(n+1)}=
\nonumber\\
&& -\gamma_1 k_BT_1k_{n,n+1}
 \int_{-\infty}^{\infty} d\omega \frac{\omega}{\pi} {\rm Im}\left[
{\rm G}^r_{n,1}(\omega)   {\rm G}^{a}_{1,n+1}(\omega)  
\right]
\nonumber\\
&&-
\gamma_N k_BT_Nk_{n,n+1}
 \int_{-\infty}^{\infty} d\omega \frac{\omega}{\pi} {\rm Im}\left[
{\rm G}^r_{n,N}(\omega)   {\rm G}^{a}_{N,n+1}(\omega)  
\right]
\nonumber\\
&&-
\gamma_1 k_BT_Nk_{n,n+1}
 \int_{-\infty}^{\infty} d\omega \frac{\omega}{\pi} {\rm Im}\left[
{\rm G}^r_{n,1} (\omega)  {\rm G}^{a}_{1,n+1(\omega)}  
\right]
\nonumber\\
&&+
\gamma_1 k_BT_Nk_{n,n+1}
 \int_{-\infty}^{\infty} d\omega \frac{\omega}{\pi} {\rm Im}\left[
{\rm G}^r_{n,1}(\omega)   {\rm G}^{a}_{1,n+1}(\omega) 
\right].
\nonumber\\
\eea
However, based on the equilibrium identity (\ref{eq:N2}), lines 2 and 3 cancel out, and we get
\bea
&&J_M^{(n+1)}=
\nonumber\\
&&\gamma_1 k_B(T_N-T_1)k_{n,n+1}
 \int_{-\infty}^{\infty} d\omega \frac{\omega}{\pi} {\rm Im}\left[
{\rm G}^r_{n,1}(\omega)   {\rm G}^{a}_{1,n+1}(\omega)  
\right].
\nonumber\\
\eea
In this form, we immediately confirm that the heat current vanishes at equilibrium and that it cannot support a diode effect.

%====================================
% Figure 11 C2 
\begin{figure*} 
\includegraphics[scale=0.45]{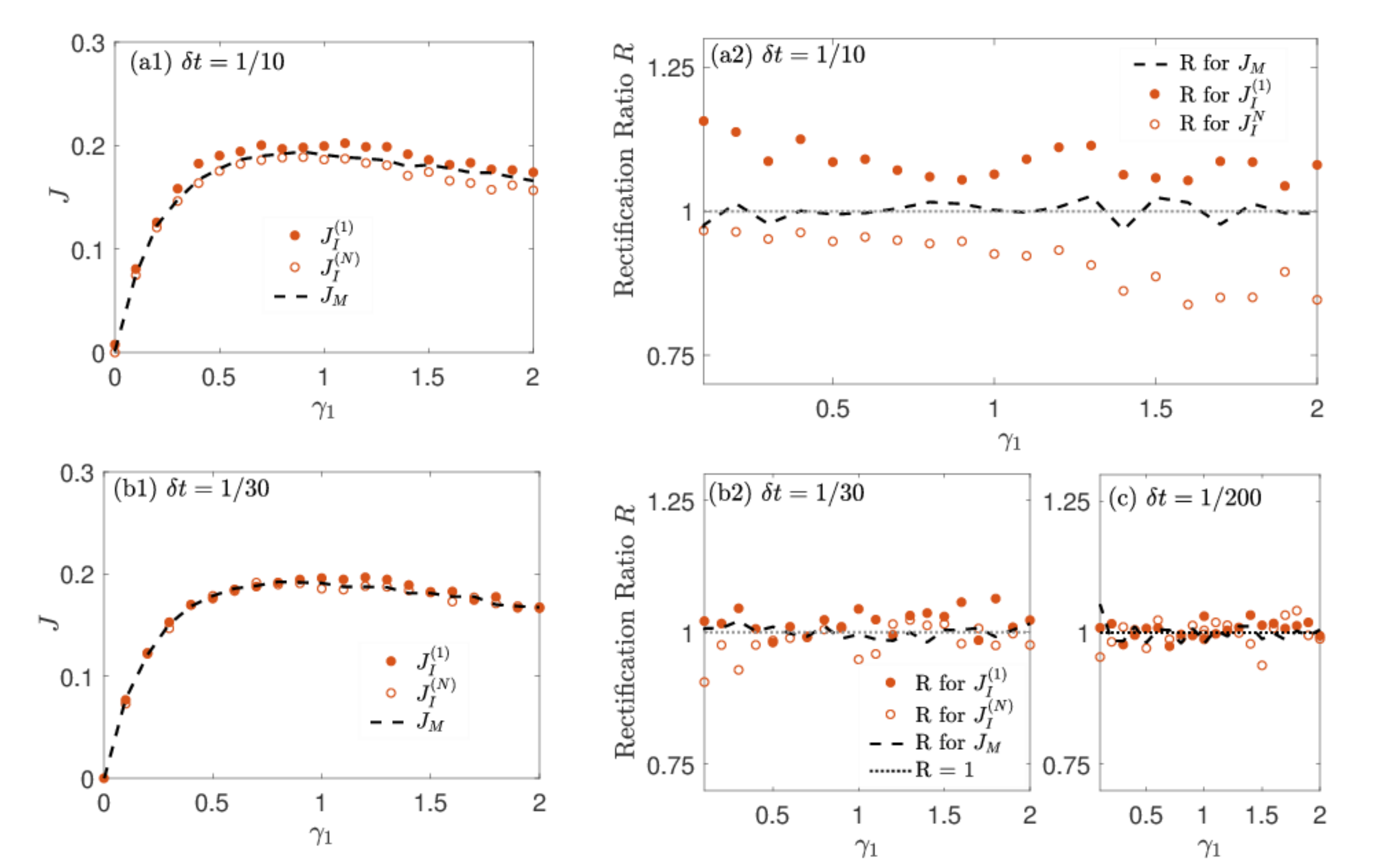}%{BBK_noneq_fric_N3.eps}
\vspace{-2mm}
\caption{Molecular dynamics simulations with a BBK scheme demonstrating the erroneous thermal rectification effect in harmonic systems for large $\delta t$.
(a1) Current and (a2) the incorrect thermal diode effect for $\delta t=0.1$.
(b1)-(b2) As we reduce the time step to $\delta t=1/30$, the interface current approaches $J_M$, and the incorrect diode effect is suppressed.
(c) Even for very short time step, $\delta t=1/200$, the intramolecular definition outperforms the interface definition, with a smaller error for $R$.
We use $N=6$, $\gamma_N=1$, $k=1$, $T_{H}=2$ and $T_C=1$.  }
\label{FigC2}
\end{figure*}
%
%% Figure 12 C3  
%\begin{figure} [htbp]
%\includegraphics[scale=0.36]{new_int_comparisonN.eps} % {dt_error_N1.eps}
%\caption{Heat currents in a nonequilibrium steady state with different integrators, RK4 (square), BBK (diagonal) and the VEC
%(triangle).
%We use $T_1=2$, $T_N=1$, $N=6$, $k=1$, $\gamma_1=\gamma_N=1$. } 
%\label{FigC3}
%\end{figure}
 %
%===============================

\subsection{Analysis of the identity (\ref{eq:N2})}
\label{sec:Iden}

Here we prove that the integral condition (\ref{eq:N2}) is identically zero for any cutoff frequency and that it does not suffer from
numerical convergence problems. This is because the integral vanishes already as a sum of two integrands.
%, unlike the normalization condition (\ref{eq:N}).
%As such, while both bulk and interface currents are equivalent to the Landauer form, the 
%bulk current is reduced to it through a trivial identity, while the interface current relies on a sum rule (\ref{eq:N}), which is difficult to converge.
%In fact the imaginary part in Eq. (\ref{eq:N2}), along with the factor of 2 in the bulk current definition 
%were introduced for symmetrization, and can be ignored in the classical case. 
We focus  on the following relation,
\bea
&&\sum_m \gamma_m 
 \int_{-\infty}^{\infty} d\omega \frac{\omega}{\pi} {\rm Im} \left[
{\rm G}^r_{n,m}(\omega)   {\rm G}^{a}_{m,n+1}(\omega)  
\right] = 0.
\nonumber\\
\label{eq:N3}
\eea 
Since only $\gamma_1$ and $\gamma_N$ are nonzero, it amounts to
\bea
&&\gamma_1 
 \int_{-\infty}^{\infty} d\omega \frac{\omega}{\pi}{\rm Im}  \left[
{\rm G}_{n,1} ^r(\omega)  {\rm G}^{a}_{1,n+1}(\omega)  
\right] 
\nonumber\\
&&+\gamma_N
 \int_{-\infty}^{\infty} d\omega \frac{\omega}{\pi} {\rm Im}  \left[
{\rm G}_{n,N} ^r(\omega)  {\rm G}^{a}_{N,n+1}  (\omega)
\right] =0.
\label{eq:N4}
\eea 
%
%
%We now show that the identity (\ref{eq:N3}) corresponds to a basic relationship for the retarded and advanced Green's functions.
 Identifying the spectral function,
\bea 
{\bf A}(\omega)& \equiv& i \big[ {\bf G}^r(\omega) - {\bf G}^a(\omega)\big] 
\nonumber\\
&=& 2 \omega {\bf G}^r(\omega){\bf \Gamma}(\omega) {\bf G}^a(\omega),
\eea
where ${\bf \Gamma}(\omega)$ is a diagonal matrix with $\gamma_n$ on the diagonal,
we note that $A_{l,l+1}(\omega)= \sum_m 2\omega \gamma_m G^r_{l,m}(\omega)G^a_{m,l+1}(\omega)$.
%Therefore, (\ref{eq:N3}) can be written as
%
%\bea
%\frac{1}{2\pi}{\rm Im}\int_{-\infty}^{\infty} d\omega  A_{l,l+1}(\omega) = 0
%\eea
 However, all the elements of ${\bf A}(\omega)$ are real: 
\begin{align}
A_{n,m} &= i[G^r_{n,m}(\omega) - G^a_{n,m}(\omega)] \nonumber \\
& = i[G^r_{n,m}(\omega) - G^{r*}_{n,m}(\omega)] \nonumber \\
&= -2\Im[G^r_{n,m}(\omega)],
\end{align}
therefore ${\rm Im }[A_{l,l+1}(\omega)] = 0$.
 
In other words, the two terms in (\ref{eq:N4}) cancel out for every frequency component,
$\gamma_1 {\rm Im} [
{\rm G}_{n,1} ^r(\omega)  {\rm G}^{a}_{1,n+1}(\omega)  ] = -\gamma_N
{\rm Im }[{\rm G}_{n,N} ^r(\omega)  {\rm G}^{a}_{N,n+1}  (\omega)]$.
The intramolecular current thus reduces to a Landauer form based on an identity that holds at the level of the integrand.
This is to be contrasted with the interface current that reduces to a Landauer form only after utilizing the
 sum-rule (\ref{eq:N}), which is satisfied at the level of the integral. 

%===============================================================
% Appendix C
\renewcommand{\theequation}{C\arabic{equation}}
\setcounter{equation}{0}  % reset counter

\section*{Appendix C: Molecular dynamics simulations with other integrators}

The BBK method is one of the most popular integrator in molecular dynamics simulations \cite{BBK}.
In Figs. \ref{FigC1}-\ref{FigC2} we display simulations that are analogous to Fig.~\ref{FigMD1} and Fig.~\ref{FigMDR}, respectively, from the main text, but generated here with the BBK integrator, rather than with RK4.
First, in Fig.~\ref{FigC1} we study the equilibrium case. The intramolecular current is very close to zero, while the interface currents  show more substantial deviations. 
%The error depends on the  friction coefficient $\gamma_1$; $\gamma_N$ is fixed here.
Fig.~\ref{FigC2}  demonstrates the unphysical thermal diode effect that shows up with the interface definition when not properly converged.

%In Fig.~\ref{FigC3} we further employ the  integrator of Venden-Eijnden and Ciccotti \cite{Ciccotti}, which is fitting for propagating stochastic differential equation with a noise that is non-continuous. While the three integrators have different time-discretization errors, they all display the same characteristics, that with reducing timestep, the intramolecular current converges more quickly to the correct limit, than the interface current. 
%It is intriguing to note that with both the BBK and VEC methods, $-J_I^{(1)}>J_M$, while the opposite holds for RK4. To explain these trends one needs to carefully study errors associated with the different integrators, and how these errors impact energy equipartition and achieving local equilibrium. 

%==========================================


\begin{thebibliography}{99}
\bibitem{ohmic}
Note that the Langevin equation corresponds to Ohmic baths  
\cite{Tuckerman} with the self energies $\Sigma_{1,N}(\omega)=i\omega\gamma_{1,N}$ \cite{DharRev}. %DDD
\end{thebibliography}
\end{document}